\documentclass[twocolumn,showpacs,showkeys,amssymb,nobibnotes,superscriptaddress,aps,prx,bibnotes]{revtex4-2}

\usepackage{graphicx}% Include figure files
\usepackage{dcolumn}% Align table columns on decimal point
\usepackage{bm}% bold math
\usepackage[T1]{fontenc}
\usepackage{mathptmx}
\usepackage{mathrsfs}
\usepackage{physics}
\usepackage{simplewick}
\usepackage{amsthm,amsmath,amssymb}

\usepackage[bottom]{footmisc}

\usepackage{amssymb}
\usepackage{amsmath}
\usepackage{amsfonts}
\usepackage{bm}
\usepackage{graphicx}
\usepackage{subfigure}
\usepackage[dvipsnames]{xcolor}
\usepackage{calc}
\usepackage{epsfig}
\usepackage{color}
\usepackage[colorlinks,citecolor=blue]{hyperref}
\begin{document}
\bibliographystyle{apsrev4-1}
\newcommand{\be}{\begin{equation}}
\newcommand{\ee}{\end{equation}}
\newcommand{\bs}{\begin{split}}
\newcommand{\es}{\end{split}}
\newcommand{\R}[1]{\textcolor{red}{#1}}
\newcommand{\B}[1]{\textcolor{blue}{#1}}
\newcommand{\YC}[1]{\textcolor{orange}{#1}}

\title{Semi-classical gravity phenomenology under the causal-conditional quantum measurement prescription}
\author{Yubao Liu}
\affiliation{Center for Gravitational Experiment, Hubei Key Laboratory of Gravitation and Quantum Physics, School of Physics, Huazhong University of Science and Technology, Wuhan, 430074, China}
\author{Haixing Miao}
\affiliation{State Key Laboratory of Low Dimensional Quantum Physics, Department of Physics, Tsinghua University, Beijing, China}
%\author{Animesh Datta}
%\affiliation{Department of Physics, University of Warwick, Coventry CV4 7AL, United Kingdom}
\author{Yanbei Chen}
\email{yanbei@caltech.edu}
\affiliation{Burke Institute of Theoretical Physics, California Institute of Technology, Pasadena, CA, 91125, USA}
\author{Yiqiu Ma}
\email{myqphy@gmail.com}
\affiliation{Center for Gravitational Experiment, Hubei Key Laboratory of Gravitation and Quantum Physics, School of Physics, Huazhong University of Science and Technology, Wuhan, 430074, China}
\affiliation{Department of Astronomy, School of Physics, Huazhong University of Science and Technology, Wuhan, 430074, China}

\begin{abstract}
The semi-classical gravity sourced by the quantum expectation value of the matter's energy-momentum tensor will change the evolution of the quantum state of matter. This effect can be described by the Schroedinger-Newton\,(SN) equation, where the semi-classical gravity contributes a gravitational potential term depending on the matter quantum state. This state-dependent potential introduces the complexity of the quantum state evolution and measurement in SN theory, which is different for different quantum measurement prescriptions. Previous theoretical investigations on the SN-theory phenomenology in the optomechanical experimental platform were carried out under the so-called post/pre-selection prescription. This work will focus on the phenomenology of SN theory under the causal-conditional prescription, which fits the standard intuition on the continuous quantum measurement process. Under the causal-conditional prescription, the quantum state of the test mass mirrors is conditionally and continuously prepared by the projection of the outgoing light field in the optomechanical system. Therefore a gravitational potential depends on the quantum trajectory is created and further affects the system evolution. In this work, we will systematically study various experimentally measurable signatures of SN theory under the causal-conditional prescription in an optomechanical system, for both the self-gravity and the mutual gravity scenarios. Comparisons between the SN phenomenology under three different prescriptions will also be carefully made. Moreover, we find that quantum measurement can induce a classical correlation between two different optical fields via classical gravity, which is difficult to be distinguished from the quantum correlation of light fields mediated by quantum gravity.
\end{abstract}

\maketitle
\section{Introduction}
Einstein's General Theory of Relativity reveals the nature of gravity as a spacetime curvature $G_{\mu\nu}$ that coupled to the matter energy-stress tensor $T_{\mu\nu}$, which can be summarised as Einstein's field equation $G_{\mu\nu}=8\pi T_{\mu\nu}$. In Einstein's theory, both spacetime geometry and matter are classical. However, the physical law that governs matter evolution is quantum mechanics, which means that the energy-stress tensor should be an operator $\hat T_{\mu\nu}$ in the quantum world. Therefore, quantising the spacetime geometry $G_{\mu\nu}$ is one natural approach to establishing a consistent description of gravity\,\cite{Belenchia2018,Wald2020}, which is yet to be successful. On the other hand, there is also an alternative semi-classical approach in which the spacetime geometry remains classical, while it is sourced by the quantum expectation of the stress-energy operator, i.e. $G_{\mu\nu}=8\pi\langle \psi|\hat T_{\mu\nu}|\psi\rangle$ ($|\psi\rangle$ is the quantum state of matter which evolves with the spacetime) as originally proposed by M{\"o}ller and Rosenfeld\,\cite{Mueller1962,Rosenfeld1963}. 

Although quantum gravity seems the most natural and logical way forward, and despite many arguments against classical gravity (in particular its inconsistency with Everett's relative-state interpretation raised by Page~\cite{Everett1957,Page1981,Anastopoulos_2014}), classical gravity has not been ruled out completely. Unlike the other three fundamental interactions, there is currently no direct experimental evidence for the quantumness of the gravitational field due to the demanding condition for such a test. Therefore, it is meaningful to test the quantumness of the gravitational field. In history, similar discussions on the quantumness of electromagnetic\,(EM) field also motivated the works on experimentally testing the so-called semi-classical EM theory\,\cite{Clauser1972,Clauser1974,Kimble1977,Marshall1963,Marshall1965,Francisca1988,Francisca1992}, which was once thought to be indistinguishable from quantum electrodynamics~\cite{Jaynes1963,Crisp1969,Stroud1970,Leiter1970}.

Recent developments in quantum optomechanics provides an experimental platform for testing physical phenomena at a new interface between quantum and gravitational physics.~\cite{Aspelmyer2012,Aspelmyer2014,Chen_2013}. This platform allows the preparation, manipulation, and characterisation of macroscopic objects near Heisenberg Uncertainty, and spans a wide range of experimental scenarios, including levitated nanospheres~\cite{Ballestero2021,Hoang2016,Yin2013}, nanomechanical oscillators~\cite{Mason2019,Rossi2018,Rossi2019,Brooks2012,Safavi2013,Purdy2017,Peterson2016,Kampel2017,Teufel2016}, membranes\,\cite{Jayich2008}, and suspended test masses~\cite{Yu2020,Aggarwal2020,Cripe2019}. 

Experimental opportunities motivate the study of the phenomenology of the semi-classical gravity of macroscopic, non-relativistic objects. In this paper, we shall focus on the Schroedinger-Newton theory~\cite{Diosi1989,Penrose1996,Diosi1998,Carlip_2008,Bahrami_2014,Bassi_2017}
\begin{equation}
    i\hbar\frac{\partial |\Psi\rangle}{\partial t} = H|\Psi\rangle + V_N(|\Psi\rangle) |\Psi\rangle
\end{equation}
in which the quantum state $|\Psi\rangle$  of a system of particles, as can be represented by the multi-particle joint wavefunction, is subject to a Newton gravitational potential $V$ that in turn depends on the state $|\Psi\rangle$. In Ref.~\cite{Yang2013,Helou2017}, the effect of $V_N$ is elaborated for a macroscopic test mass (a solid), when its Center-Of-Mass (COM) position uncertainty is less than the zero-point position uncertainty of atoms near their lattice sites. In phase space, the quantum uncertainty (in terms of the position-momentum covariance matrix) of the COM evolves at a shifted frequency from that of the expectation values. Further works discussed possible experimental signatures of the SN theory~\cite{Yang2013,Helou2017,Grossardt2016,Gan2016}.

%Among them, one particularly interesting work was developed by Yang \emph{et. al}\,\cite{Yang2013,Helou2017}, which shows a nonlinear modification of the Schroedinger equation of the centre-of-mass degree of freedom of a macroscopic object in the semi-classical gravity, which is called the Schroedinger-Newton theory\,. Various works has been devoted to discuss the experimental test proposals\,

%This nonlinearity can be attributed to the following reason: the gravitational potential term in the Hamiltonian is determined by the quantum expectation value of the source matter $\langle\psi|\hat T_{\mu\nu}|\psi\rangle$, therefore it must depend on the source matter's wave function\,\cite{Bassi_2017}. 

The nonlinearity in $V(|\Psi\rangle)$ brings an ambiguity to SN theory when describing quantum measurements, and in this process breaking  the \emph{time-reversal symmetry} of  standard quantum mechanics. Let us consider a scattering amplitude problem: suppose we prepare a system at an initial state $|i\rangle$, let it evolve for duration $T$, and would like to compute the probability that it will be found at $|f\rangle$.  Even though the SN equation {\it appears to be} time-reversal symmetric, since in general, $|i\rangle$ does not evolve into $|f\rangle$, when inserting the $|\Psi\rangle$ in $V$, one needs to specify whether to make $|\Psi\rangle$ agree with $|i\rangle$ at the initial time, or to make $|\Psi\rangle$ agree with $|f\rangle$ at the final time.   In other words, if we denote with $\hat U$ the evolution operator, then the relation $p_{i\rightarrow f}=|\langle f|\hat U|i\rangle|^2=|\langle i|\hat U|f\rangle|^2$ will not hold since $\hat U$ depends on the 
quantum state due to the nonlinearity: $\hat U(t,t_0)=\hat U_{|\psi\rangle}(t,t_0)$\,\cite{Helou2017}. In standard quantum mechanics, the wavefunction collapse has three different, but equivalent prescriptions/interpretations: pre-selection, post-selection, and conditional collapse, as discussed by Aharonov \emph{et al.} in a milstone paper\,\cite{Aharonov1964,Reznik1995}. However, the equivalence of these prescriptions depends on the linearity of quantum mechanics, that is, such an equivalence will be broken in SN theory.  For example, for post/pre-selection prescription, we have $p^{\rm pre/post}_{i\rightarrow f}=|\langle f|\hat U_{|\psi\rangle_{i/f}}|i\rangle|^2$ and clearly $p^{\rm pre}_{i\rightarrow f}\neq p^{\rm post}_{i\rightarrow f}$. 

In this paper, we will introduce a version of the so-called {\it causal-conditional prescription} of SN theory~\cite{Bassam2017b}, which differs from the pre/post-selection prescriptions, leading to significantly different phenomenology.  In the non-relativistic limit of this prescription, Newton's potential will be determined by the instantaneous conditional quantum state of the system. 
%
%For example, suppose we have 
Consider a system of two macroscopic mirrors (A and B), interacting via their \emph{mutual gravitational interaction}, with each of the centre-of-mass position $x_{A}$ and $x_B$  monitored by a separate optical field. In previous theoretical proposals~\cite{Miao2020,Carney2022}, it was assumed that, under SN theory, the motion of mirror A, driven by {\it quantum radiation-pressure fluctuations}, will have a vanishing quantum expectation, therefore will not drive the motion of mirror B.   This argument predicts zero correlation between the two out-going optical fields $y_A$ and $y_B$. In this way, any correlation between the two output fields can be used to verify the quantum nature of gravity.

%Such a system has been discussed before in the quantum gravity case and predicted that there is no correlation between the two output optical fields in the semi-classical gravity\,(SN) case\,. 

%
However, as we shall see in this paper, under the causal-conditional prescription, the \emph{continuous quantum measurement} of mirror motions can actually induce correlations between the two out-going fields.  We can in fact argue that some degrees of correlation between the two optical fields must exist in SN theory, in which the quantum state $|\Psi\rangle$ used in generating Newton's potential is {\it updated} according to measurement results. In the specific case of the two mirrors under causal-conditional prescription, the conditional state of mirror A, hence the conditional expectation of $E( x_A|y_A)$, evolves in a way that depends on measurement results $y_A$, hence A exerts a classical gravitational force on B that is correlated with $y_A$, which in turn establishes a correlation between $y_A$ and $y_B$. As it turns out, in comparison with zero correlations, the \emph{classical correlation} predicted by the causal conditional prescription  is much more difficult to distinguish from the quantum-gravity-induced correlations of light fields when the two mirrors are interacting via weak quantum gravity.  This result provides an important  caution toward experimental verification of the quantum nature of gravity. 

This paper is structured as follows. Section\,\ref{sec.2} will give a general discussion of the continuous quantum measurement under different prescriptions for SN theory. Then in Section\,\ref{sec.3} and\,\ref{sec.4}, the optomechanical systems with the strong SN effect\,(semi-classical self-gravity scenario) and the weak SN effect\,(semi-classical mutual-gravity scenario) will be thoroughly analysed, respectively. Section\,\ref{sec.5} will discuss the physical origin of the semi-classical gravity-induced light field correlations from the aspect of non-linear quantum mechanics. Finally, in Section\,\ref{sec.6} summary and discussions of this work will be presented.

\section{Continuous quantum measurement in SN theory}\label{sec.2}
The first step to study the SN phenomenology is to establish a theoretical description of the continuous quantum measurement in SN theory. As we have mentioned in the \emph{Introduction}, the quantum state evolution and measurement induced state collapse in a non-linear quantum mechanical theory are different from the standard quantum mechanics. This is because \emph{the symmetry of pre and post-selection prescriptions in the standard quantum mechanics is no longer valid}, which has been extensively discussed and applied to the analysis of the measurement of a single test mass mirror exerted by its self-gravity in\,\cite{Yang2013} with the following Hamiltonian:
\be\label{eq:Hamiltonian_self_gravity}
\hat{H}=\frac{\hat{p}^2}{2M}+\frac{1}{2}M\omega_{\rm{m}}^2\hat{x}^2+\frac{1}{2}M\omega_{\rm{SN}}^2(\hat{x}-\langle\hat{x}\rangle)^2-\hbar\alpha\hat{a}_1\hat{x},
\ee
where $\alpha$ is the measurement strength that proportional to the coherent amplitude of the pumping light and $\hat a_1$ is the amplitude operator of the optical flucutations. The $\omega_{\rm SN}=\sqrt{Gm/6\sqrt{\pi}x_{\rm zp}^3}$ is the SN frequency\,\cite{Yang2013}, in which the $m,x_{\rm zp}$ are the mass and zero-point displacement of the crystal lattice oscillation in the test mass mirror.

In the cases of the pre/post-selection (as discussed in\,\cite{Helou2017}), the evolution operator $\hat U_{|\psi\rangle}={\rm exp}[-i\int_0^t\hat H(|\psi_m(t)\rangle)dt]$ depends on the initial/final mechanical quantum state. This means that $\langle \hat x\rangle$ in the above Hamiltonian can be treated as $\langle \psi_m(t_{i/f})|\hat x(t)|\psi_m(t_{i/f})\rangle$ throughout the entire continuous measurement processes, which is a deterministic c-number. Therefore this approach linearised the problem. 

\begin{figure}[h]
\centering
\includegraphics[width=0.4\textwidth]{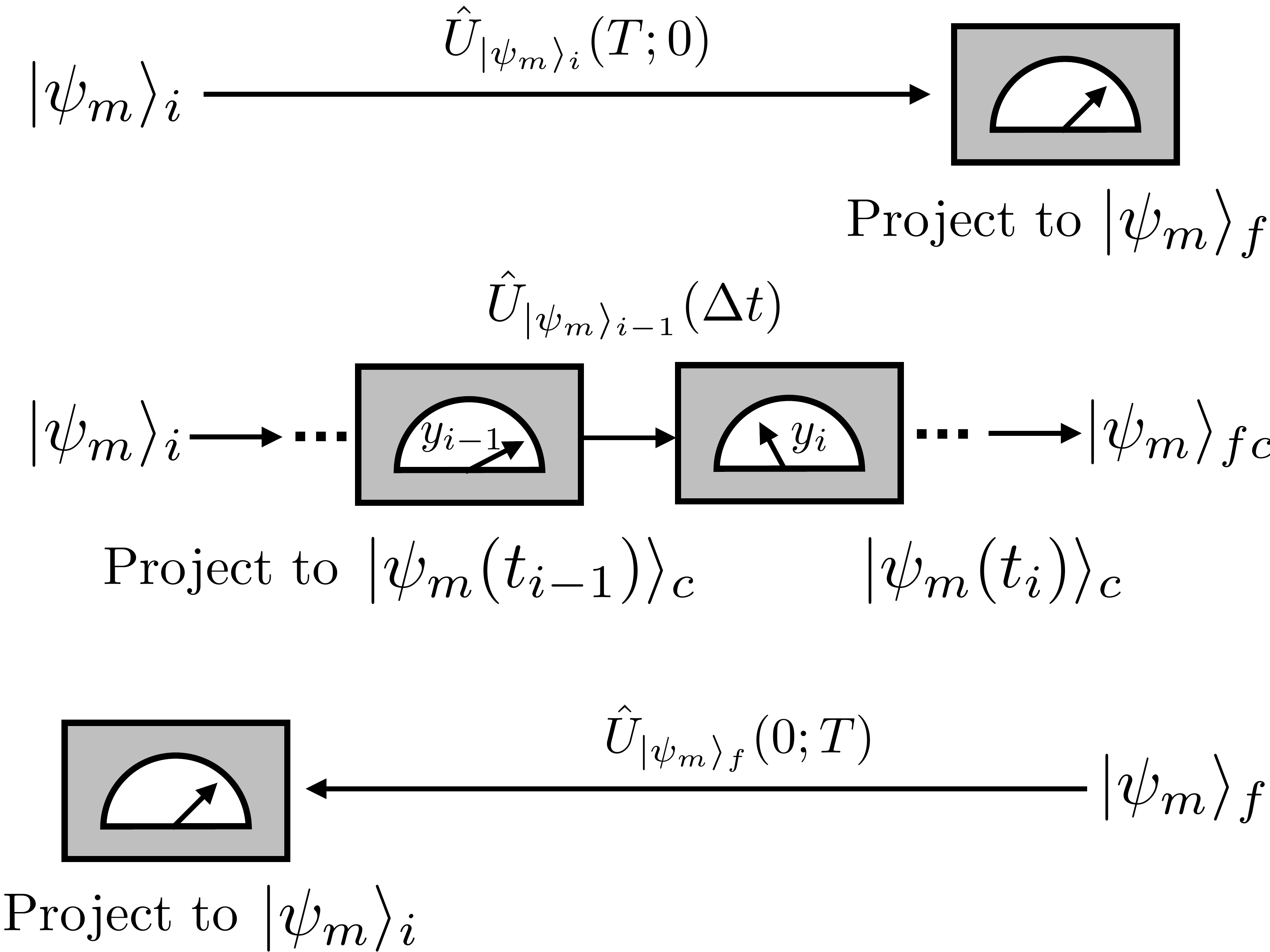}
\includegraphics[width=0.45\textwidth]{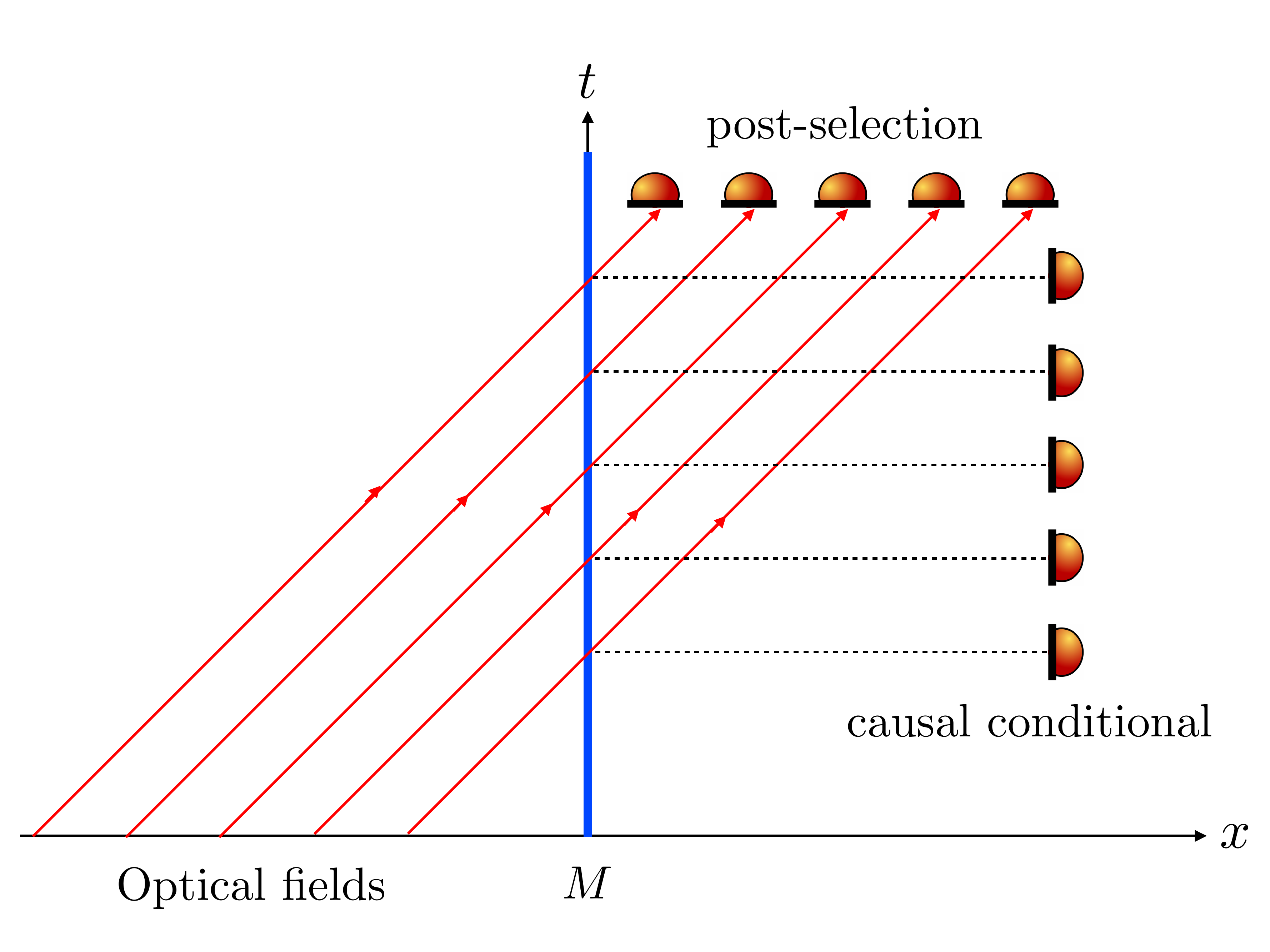}
\caption{Upper panel: Different prescription of state evolution and measurement in the Schroedinger-Newton theory: (a) the post-selection prescription; (b) the causal-conditional prescription where the measurement-evolution processes satisfy Eq.\,\eqref{eq:causal_conditional_scheme}, where the conditional mechanical quantum state prepared continuously follows a quantum trajectory; (c) the pre-selection prescription. Lower panel: Spacetime diagram for quantum measurement, where the thick blue line represents the worldline of the mirror.}\label{fig:prescription}
\end{figure}

However, for continuous measurement, a more intuitive prescription is the \emph{casual conditional} prescription, which can be represented by:
\be\label{eq:causal_conditional_scheme}
|\psi_m(t_f)\rangle=\hat U_{|\psi_{N-1}\rangle}^{\Delta t}\hat P_{N-1}\hat U_{|\psi_{N-1}\rangle}^{\Delta t}...\hat P_1\hat U_{|\psi_0\rangle}^{\Delta t}|0\rangle_o\otimes|\psi_m(t_i)\rangle,
\ee 
where the $\hat U_{|\psi_{j}\rangle}^{\Delta t}$ is the SN evolution of quantum state in a infinitestimal duration $\Delta t$ when the mechanical state is $|\psi_{j}\rangle$, the $\hat P_j$ is the projection operator acting on the light field at time $t_0+j\Delta t$. The projective measurement on the light field will prepare the joint entangled opto-mechanical state $\hat U_{|\psi_{j}\rangle}^{\Delta t}|\psi_{j}\rangle\otimes|0\rangle_j$ onto a conditional mechanical quantum state $|\psi_{j+1}\rangle$ with measurement record $y_i$. This causal-conditional prescription is equivalent to the pre/post selection prescription only in standard quantum mechanics. This inequivalency can be seen from the fact that in the SN theory $\hat U_{|\psi_{N}\rangle}^{\Delta t}...\hat U_{|\psi_{0}\rangle}^{\Delta t}\neq \hat U_{|\psi_{0}\rangle}(t_N,t_0)$.

A direct result of the causal-conditional prescription is the dependence of the gravitational field on the stochastic quantum trajectory\,\cite{Doherty1999} since the gravitational field in the SN theory is sourced by the conditional quantum expectations of the mirror's physical quantities. In contrast, under the pre/post-selection prescription, the gravitational field throughout the continuous quantum measurement process has a deterministic evolution, which will exhibit a different phenomenology. Interestingly, the gravitational field evolution under the causal-conditional prescription is somewhat similar to the quantum gravity, where the gravitational field is sourced by the mirror exerted by the stochastic quantum radiation pressure noises. These points will be elaborated in the following sections using the optomechanical systems as an example.

Optomechanical system is the most promising experimental platform for testing macroscopic quantum mechanics\,\cite{Chen_2013,Ebhardt2009,Schnabel2015,Mason2019,Rossi2018,Matsumoto2020}. In the following, we will give a complete analysis on the phenomenology of semi-classical gravity on the optomechanical system, in particular under the causal-conditional prescription.  We will discuss two different scenarios: (1) the optomechanical system influenced by the mirror's the semi-classical \emph{self-gravity}, where the SN effect is relatively strong since the gravity interaction happens at length scale $\sim x_{\rm zp}$; (2) the optomechanical system with two mirrors interacting via \emph{mutual semi-classical gravity}, where the SN effect is relatively weak since the gravity interaction happens at the mirror separation length scale.

\section{Optomechanical system influenced by self-gravity}\label{sec.3}
\subsection{Results of the Pre/post-selection prescription: an overview}
For an optomechanical quantum measurement system, the signatures of SN theory can be studied through the traditional preparation-evolution-verification process sketched in\,\cite{Yang2013}. In this scenario, the mirror is first prepared onto a mechanical squeezed state, then undergoes an SN free evolution and finally performs quantum tomography on the evolved state. The signature of the SN effect manifests itself in the free evolution stage, where the (Gaussian) Wigner function of the squeezed mechanical state rotates in the phase space around its mean value $(\langle \hat x\rangle_c,\langle\hat p\rangle_c)$ at frequency $\omega_q=\sqrt{\omega_m^2+\omega_{\rm SN}^2}$. The scenario this work focus on is similar to that discussed in\,\cite{Helou2017}, where we directly measure the spectrum of the outgoing optical field that interacts with the quantum test mass mirror exerted by the semi-classical gravitational field described by the SN theory. 

\begin{figure}[h]
\centering
\includegraphics[width=0.4\textwidth]{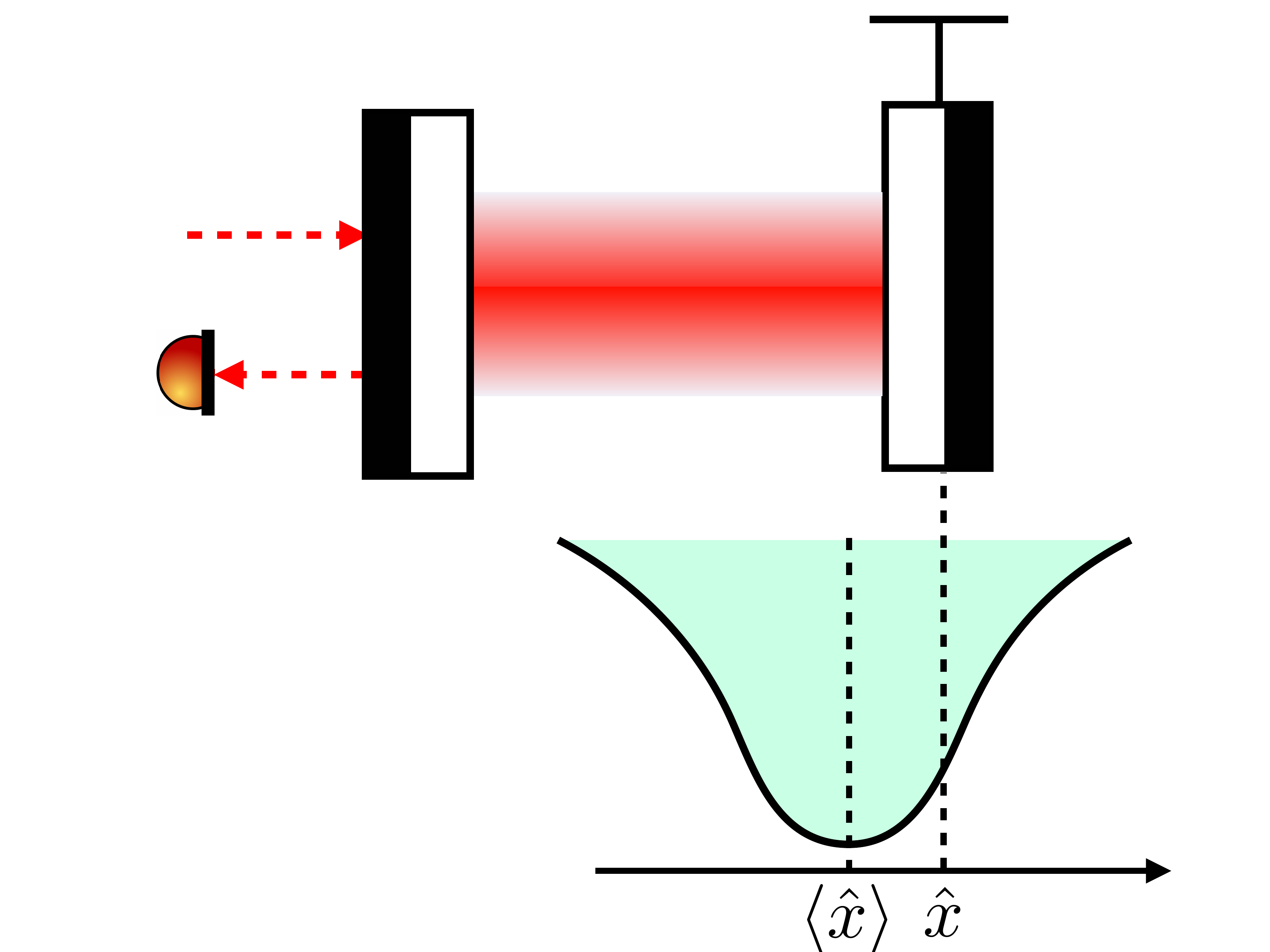}
\caption{A single cavity optomechanical system with semi-classical self-gravity. The mirror is in the classical gravitational potential created by the quantum expectation value of its stress-energy tensor, thereby having a Schroedinger-Newton correction to its pendulum frequency. This effect does not exist when gravity follows quantum mechanical law.}\label{fig:schematic_setup}
\end{figure}

In\,\cite{Helou2017}, the signatures of the SN theory are studied under the pre/post-selection prescription, which can be  summarised as a Lorentzian peak in the spectrum of the outgoing field around $\omega_q$ as:
\be
\Delta S_{a_2a_2}(\omega)\approx \beta(\beta+2)\frac{\gamma_m^2}{\gamma_m^2+4(\omega-\omega_q)^2},\quad \beta=\frac{\alpha^2}{M\hbar\gamma_m\omega_q},
\ee
where a high $Q-$oscillator is assumed, i.e $\gamma_m\ll\omega_q$.

\begin{figure}[h]
\centering
\includegraphics[width=0.4\textwidth]{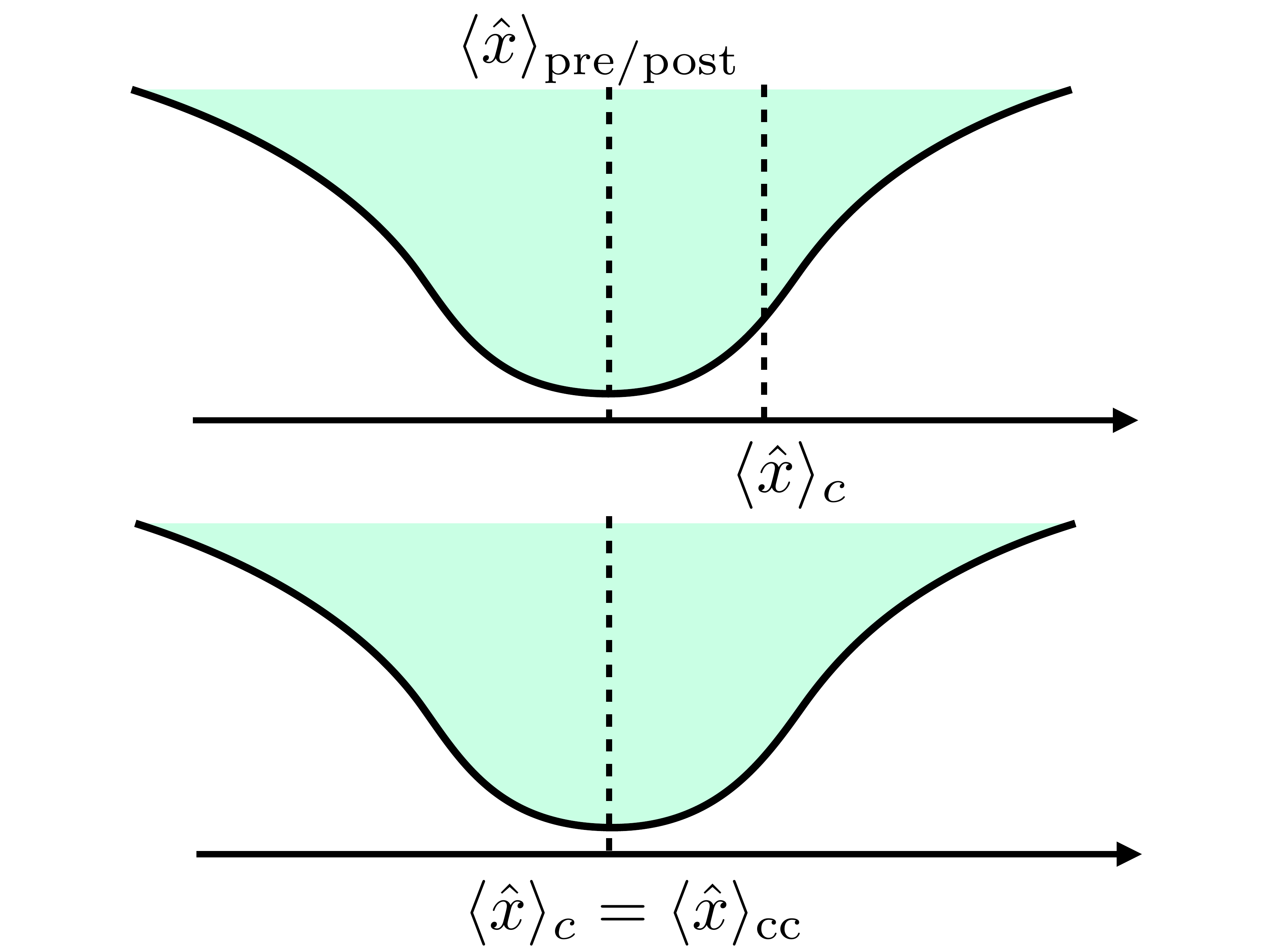}
\caption{Physical picture of the phenomenology under different prescriptions. Upper panel: in the pre/post-selection prescriptions, the gravitational potential is contributed by a fixed source with $\langle \hat x\rangle_{\rm pre/post}=\langle \psi_m(t_{i/f})|\hat x(t)|\psi_m(t_{i/f})\rangle$, while the conditional quantum expectation of the mirror position during the measurement process $\langle \hat x\rangle_c\neq \langle \hat x\rangle_{\rm pre/post}$, which contributes to a restoring force.
Lower panel: under the causal-conditional prescription, the $\langle \hat x\rangle_c$ always locates at the potential minimum, which doesn't feel a restoring force.  }\label{fig:comparison_phys}
\end{figure}

While for the post-selection prescription, the signature of the SN theory is, on the contrary, a \emph{Lorentzian dip} in the spectrum of the outgoing field around $\omega_q$:
\be
\Delta S_{a_2a_2}(\omega)\approx-\beta(\beta+2)\frac{\gamma_m^2}{(1+\beta)^2\gamma_m^2+4(\omega-\omega_q)^2}.
\ee
Besides, the outgoing field spectrum has another peak at around $\omega_m$.  The SN observational feature at around $\omega_q$ is because the conditional mean position of the mechanical quantum state under the continuous quantum measurement does not coincide with the $\langle \hat x\rangle_{\rm pre/post}$ under the pre/post selection prescription. This means that during the quantum measurement process, the conditional quantum expectation value of mechanical state feels a restoring force $\propto -m\omega^2_{\rm SN}(\langle \hat x\rangle_{\rm pre/post}-\langle \hat x\rangle_{c})$ (see Fig.\,\ref{fig:comparison_phys}).  

As we shall see in the next section, the outgoing field spectrum in the case of the causal-conditional prescription will be different: the peak/dip around $\omega_q$ does not exist. We plot the comparison of the outgoing field spectrum of these three different quantum measurement prescriptions in Fig.\,\ref{fig:spectrum}. The numerical analysis is based on the sampling parameters listed in Tab.\,\ref{tab:self_gravity}.

\begin{table}[h!]
    \centering
    \begin{tabular}{|c|c|c|}
    \hline
Parameters&Symbol&Value\\
\hline
Mirror mass&$M$&0.2\,kg\\
\hline
Mirror bare frequency&$\omega_m/(2\pi)$&$4\times10^{-3}\,{\rm Hz}$\\
\hline
SN frequency&$\omega_{\rm SN}/(2\pi)$&$7.8\times10^{-2}\,{\rm Hz}$\\
\hline
Quality factor&$Q_m$&$10^7$\\
\hline
Mechanical damping&$\gamma_m/(2\pi)$&$4\times10^{-10}\,{\rm Hz}$\\
\hline
Optical wavelength&$\lambda$&$1064\,{\rm nm}$\\
\hline
Cavity Finesse&$\mathcal{F}$&$300$\\
\hline
%$\alpha$&$4.5\times10^{14}\sqrt{{\rm Hz}}/m$\\
Intra-cavity power&$P_{\rm cav}$&$480\,{\rm nW}$\\
\hline
    \end{tabular}
    \caption{The parameters of the optomechanical system with a single mirror exerted by its semi-classical self-gravity.}
    \label{tab:self_gravity}
\end{table}

\subsection{Causal-conditional prescription for the semi-classical self-gravity}

\subsubsection{Output optical spectrum}
Following the above causal-conditional prescription, we can obtain the following stochastic master equation (SME) for describing the evolution of conditional mechanical state under continuous quantum measurement in the SN theory. The projective measurement result of the optical quadrature fields at homodyne angle $\theta$ is given as $\hat y_\theta=1/\sqrt{\Delta t}\int_{t}^{t+\Delta t}\hat{a}_\theta(t')dt'$, which satisfies: $y_\theta=\alpha\langle \hat{x}\rangle\sin\theta\sqrt{\Delta t}+\Delta W/\sqrt{2\Delta t}$.
For later use, we redefine $\tilde{a}_\theta$ as $\hat y_\theta/\sqrt{\Delta t}$ thereby:
\be\label{eq:a2}
\tilde{a}_\theta=\alpha\langle\hat{x}\rangle_c\sin\theta+dW/{\sqrt{2}dt}.
\ee
The corresponding stochastic master equation is:
\be\label{eq:sme}
\begin{split}
d\hat{\rho}=&-\frac{i}{\hbar}[\hat{H}_0,\hat{\rho}]dt-\frac{\alpha^2}{4}[\hat{x},[\hat{x},\hat{\rho}]]dt-\frac{i\alpha}{\sqrt{2}}\cos\theta[\hat x,\hat \rho]dW\\
&+\frac{\alpha}{\sqrt{2}}\sin\theta\{\hat{x}-\langle\hat{x}\rangle,\hat{\rho}\}dW-\frac{i\gamma_m}{2\hbar}\left\{\hat{x}, \{\hat{p},\hat{\rho}\}\right\}dt,
\end{split}
\ee
where $\hat H_0=\hat H+\hat H_{\rm SN}$ is the free mechanical Hamiltonian (including its self-gravitational interaction) in the SN theory, the second and third term on the right hand side (r.h.s) is the standard Lindblad term and the Ito-term, respectively. The last term describes the mechanical thermal dissipation. This mechanical dissipation term is important since the system can not reach a stationary stochastic process without it. Physically it is due to the fact that the interaction of light field with the mechanical motion (when the pumping field is on-resonance with the cavity) can only re-distribute quantum information without dynamical energy exchange.

Using Eq.\,\eqref{eq:sme}, the conditional expectation and variance of $\hat{x}$ and $\hat{p}$ are respectively given as:
\be\label{eq:expectation}
\begin{split}
&d\langle\hat{x}\rangle_c=\frac{\langle\hat{p}\rangle_c}{M}dt+\sqrt{2}\alpha V^c_{xx}\sin\theta dW,\\
&d\langle\hat{p}\rangle_c=-M\omega_{\rm{m}}^2\langle x\rangle_c dt-\gamma_m\langle\hat{p}\rangle_c dt+\sqrt{2}\alpha V^c_{xp}\sin\theta dW\\
&\qquad\qquad\qquad+\frac{\hbar\alpha}{\sqrt{2}}\cos\theta dW,
\end{split}
\ee
and
\begin{equation}\label{eq:variance}
\begin{split}
&\dot{V}^c_{xx}=\frac{2V^c_{xp}}{M}-2\alpha^2\sin^2\theta V_{xx}^{c2},\\
&\dot{V}^c_{xp}=\frac{V^c_{pp}}{M}+M\omega_q^2V^c_{xx}-2\alpha^2\sin^2\theta V^c_{xx}V^c_{xp}-\alpha^2\sin\theta\cos\theta\hbar V_{xx},\\
&\dot{V}^c_{pp}=-2M\omega_q^2V^c_{xp}-2\alpha^2\sin^2\theta V_{xp}^{c2}-2\alpha^2\sin\theta\cos\theta\hbar V_{xp}\\&\qquad\qquad\qquad-\frac{\alpha^2\cos^2\theta\hbar^2}{2}+\frac{1}{2}\alpha^2\hbar^2,
\end{split}
\end{equation}
where $\omega_q\equiv\sqrt{\omega_{\rm m}^2+\omega_{\rm{SN}}^2}$. 

It is important to note that the oscillation frequency of the conditional expectation value of the mechanical displacement $\langle \hat x\rangle_c$ is $\omega_m$ rather than $\omega_q$ under the causal-conditional prescription, which is different from the pre/post-selection prescription. This can be understood from the fact that the gravitational field at time $t$ in this case is sourced by the conditional mechanical state $|\psi_m(t)\rangle_{c}$, which also follows a random trajectory. The $\langle x \rangle_c$ in this case is always located at the potential minimum thereby feeling no restoring force. In contrast, the gravitational field under the pre/post-selection prescriptions is sourced by the deterministic evolving mechanical state $\hat U_{|\psi_m(t_{i/f})\rangle}|\psi_m(t_{i/f})\rangle$, therefore feels a restoring force as discussed in the previous subsection. This is an important difference, which will change the features of the outgoing light spectrum.

With the above equations, the conditional mean displacement can be formally solved as (assuming that the phase quadrature is measured $\theta=\pi/2$):
\begin{equation}\label{eq:xtrajectory}
\begin{split}
\langle\hat{x}(t)&\rangle_c=e^{-\gamma_m t/2}x^{(0)}(t)+\sqrt{2}\alpha\int_0^t dW(s)e^{-\gamma_m(t-s)/2}\\
&\times\left[V^c_{xx}\cos{\omega_{\rm m}(t-s)}+\frac{V^c_{xp}}{M\omega_{\rm m}} \sin{\omega_{\rm m}(t-s)}\right.\\
&+\left.\left(\frac{V^c_{xp}}{4M\omega_{\rm m} Q_m^2}+\frac{V^c_{xx}}{2Q_m}\right) \sin{\omega_{\rm m}(t-s)}\right],
\end{split}
\end{equation}
where $x^{(0)}(t)=x_0\cos(\omega_{\rm m}t)+(p_0/M)\sin(\omega_{\rm m}t)$ is the free mechanical motion which will be forgotten for $t\gg1/\gamma_m$, the terms that $\propto (1/Q_m, 1/Q_m^2)$ relates to the mechanical dissipation while the second line is purely contributed from the quantum measurement process. 

When the system reaches the steady state, the steady solution of Eq.\,\eqref{eq:variance}:
\begin{equation}\label{eq:steadyvariance}
\begin{split}
&V^c_{xx}=\frac{\hbar}{\sqrt{2}M\omega_q}\frac{1}{\sqrt{1+\sqrt{1+\Lambda^4}}},\\
&V^c_{xp}=V^c_{px}=\frac{\hbar}{2}\frac{\Lambda^2}{1+\sqrt{1+\Lambda^4}},\\
&V^c_{pp}=\frac{\hbar M\omega_q}{\sqrt{2}}\frac{\sqrt{1+\Lambda^4}}{\sqrt{1+\sqrt{1+\Lambda^4}}},
\end{split}
\end{equation}
with $\Lambda=\sqrt{\hbar\alpha^2/(M\omega_q^2)}$. These formulae give us a glimpse on the experimental requirements to distinguish quantum gravity and semi-classical gravity. In case of quantum gravity, all the $\omega_q$ in the above formula should be replaced by $\omega_m$, therefore taking $V^c_{xx}$ as an example:
\be
\frac{V^{c\rm SN}_{xx}}{V^{c\rm QG}_{xx}}=\frac{\omega_m}{\omega_q}\left[\frac{1+\sqrt{1+\Lambda_{\rm QG}^4}}{1+\sqrt{1+\Lambda_{\rm SN}^4}}\right]^{1/2},
\ee
where we redefine $\Lambda_{\rm SN/QG}=\sqrt{\hbar\alpha^2/(M\omega_{q/m}^2)}$. It is difficult to test the quantumness of gravity when the optomechanical interaction is very strong: $\Lambda_{\rm QG/SN}\gg1$, since we have $V^{c\rm SN}_{xx}/V^{c\rm QG}_{xx}\approx \omega_m\Lambda_{\rm QG}/\omega_q\Lambda_{\rm SN}=1$. This target could be achieved only when the $\Lambda_{\rm SN/QG}$ takes a moderate value and $\omega_m/\omega_q$ is not close to one. The moderate  $\Lambda_{\rm SN/QG}$ indicates that the optomechanical interaction can not be too strong, thereby low-temperature technology is required to suppress the thermal environmental effect.

\begin{figure}[h]
\centering
\includegraphics[width=0.48\textwidth]{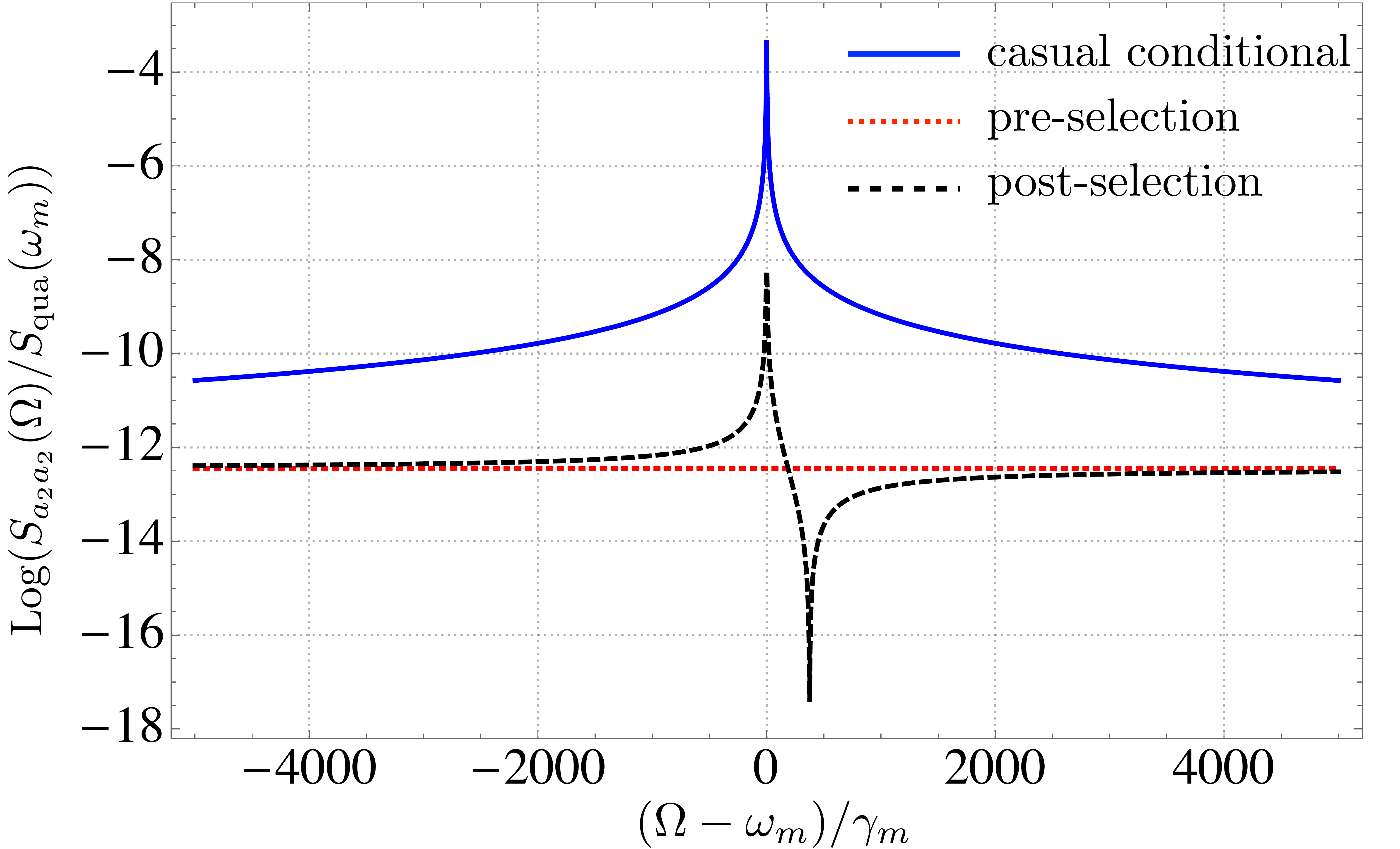}
\includegraphics[width=0.48\textwidth]{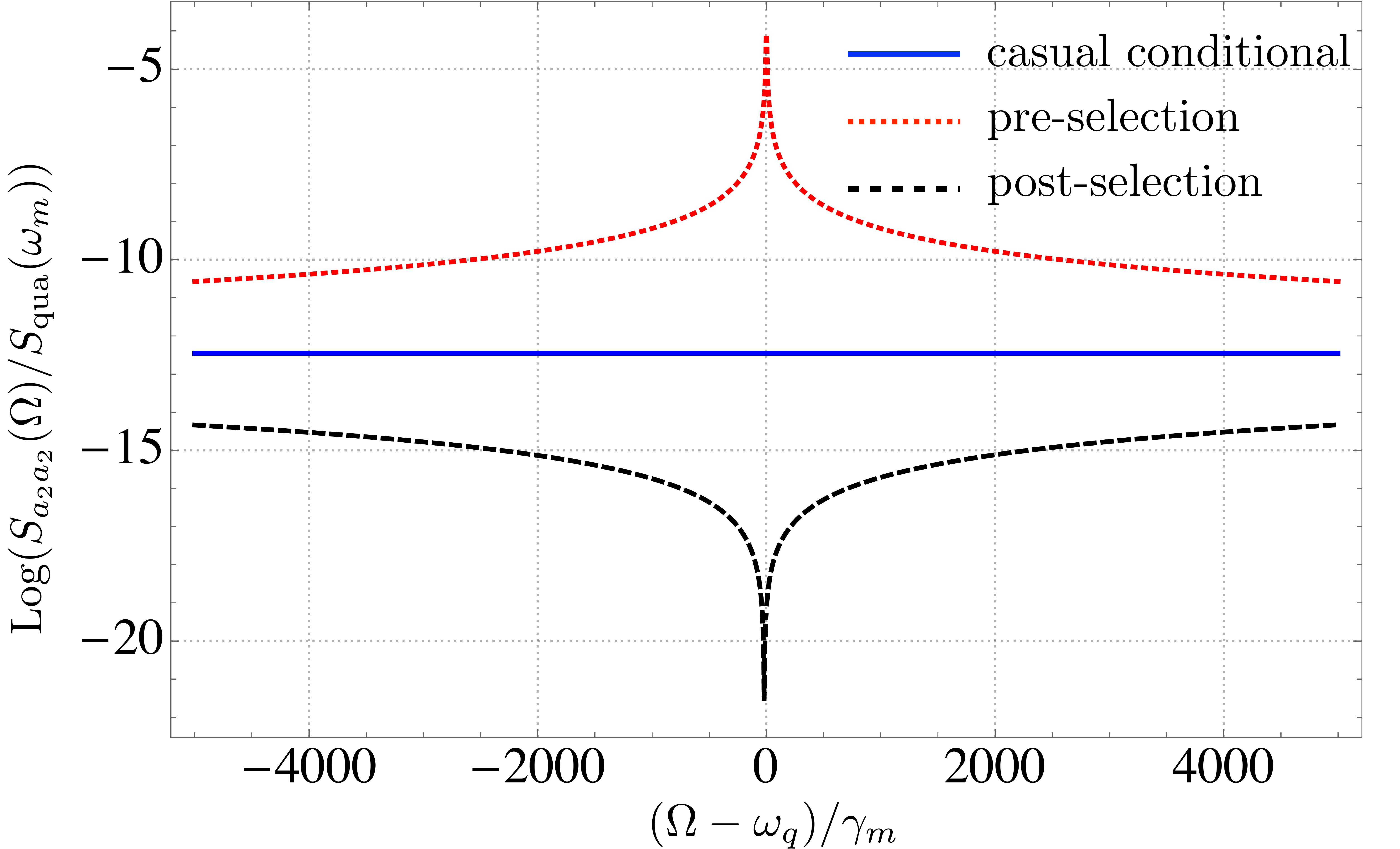}
\includegraphics[width=0.47\textwidth]{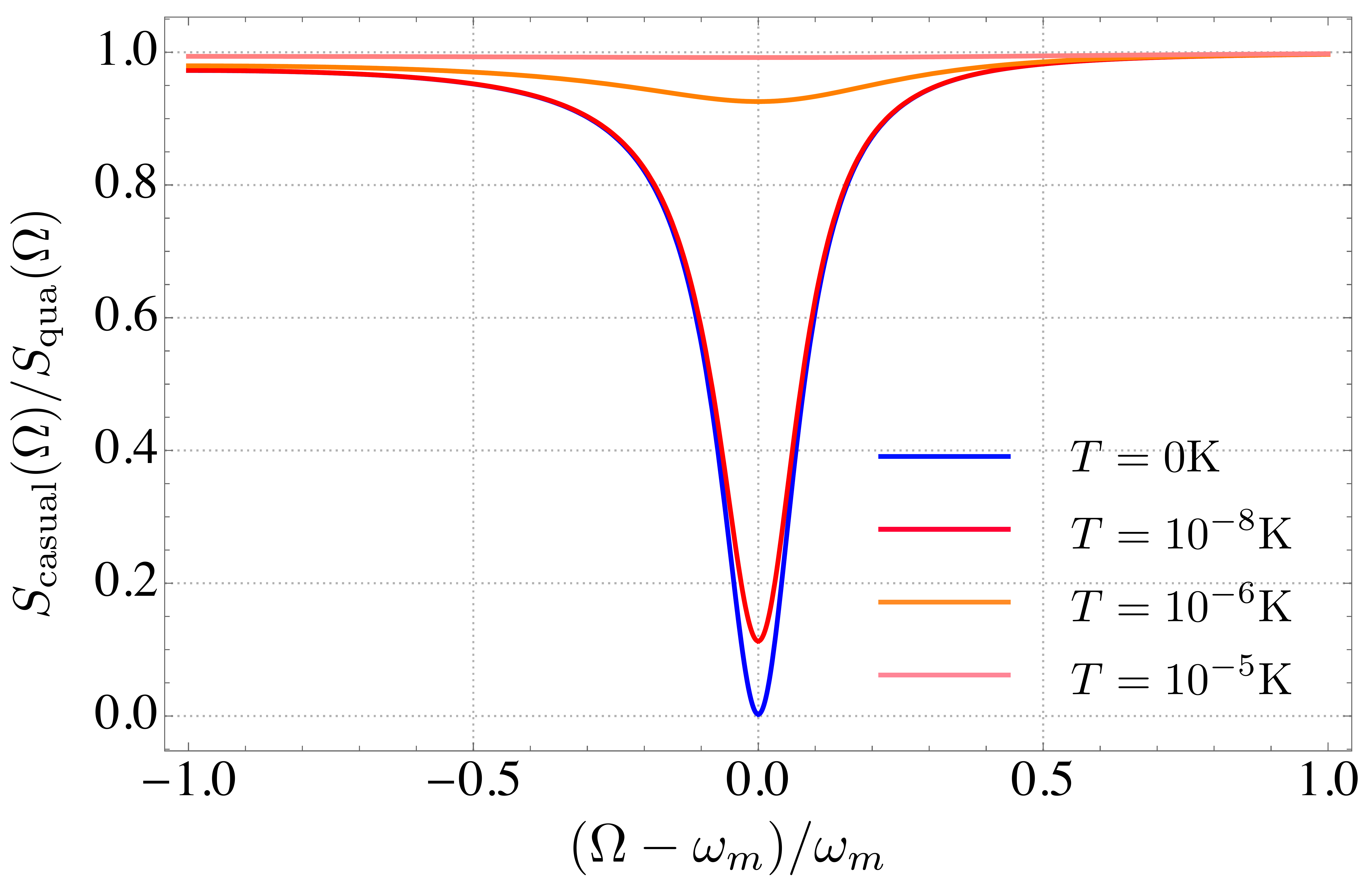}
\caption{The spectrum of the outgoing field $S_{a_2a_2}$ (quantum noise only) in the Schroedinger-Newton theory for different prescriptions around original mechanical frequency $\omega_m$ and the SN-modified frequency $\omega_q=\sqrt{\omega_m^2+\omega_{\rm SN}^2}$. All these spectrums are normalised by the standard quantum mechanics one. It is clear that the causal-conditional approach has a feature at the $\omega_m$ rather than $\omega_q$. The lowest panel shows the ratio between the outgoing spectrum of SN theory under the causal conditional prescription and the quantum gravity, where the thermal noise is considered.}\label{fig:spectrum}
\end{figure}

Combining Eqs.\,\eqref{eq:a2}\,\eqref{eq:xtrajectory} and \,\eqref{eq:steadyvariance}, we can compute the auto correlation function of $\langle\tilde{a}_2(t)\tilde{a}_2(t+\tau)\rangle$ and moreover the power density spectrum. After some tedious but straightforward algebra, the result in the high-$Q$ limit is:
\be
\begin{split}
&S_{a_2a_2}(\Omega)=\frac{4}{\gamma_m^2+4(\Omega-\omega_m)^2}\times\\
&\quad\left[\alpha^2\frac{V_{xp}}{M}\left(1-\frac{\Omega^2}{\omega_m^2}\right)+\alpha^4\left(\frac{V^2_{xp}}{M^2\omega_m^2}+\frac{\Omega^2}{\omega_m^2}V_{xx}^2\right)\right]+1,
\end{split}
\ee
where only the quantum noise is given here for illustrative purposes. It is clear that there is no quantum radiation pressure noise-induced pole around $\omega_q$ in this case, which is different from the pre/post-selection scenario where the SN signature appears around $\omega_q$ due to the reason discussed before. At the resonance point $\Omega=\omega_m$, the difference between the SN spectrum and QG spectrum has a simple formula:
\be\label{eq:self_gravity_approx}
\frac{\Delta S^{\rm SN-QG}_{a_2a_2}(\omega_m)}{ S^{\rm QG}_{a_2a_2}(\omega_m)}=\frac{2}{\sqrt{1+\sqrt{1+\Lambda_{\rm SN}^4}}}\frac{\Lambda_{\rm QG}^4Q^2}{1+\Lambda_{\rm QG}^4Q^2}\left(\frac{\omega_{\rm SN}^2}{\omega_q^2}\right)
\ee
For a strong optomechanical coupling, we have $\Lambda_{\rm QG/SN}\gg1$ which suppressed the difference. For a weak optomechanical coupling with $\Lambda_{\rm SN/QG}\sim 1$, the difference could be significant for a high Q oscillator when we ignore the thermal noise as shown in the third panel of Fig.\,\ref{fig:spectrum}. However, this difference will be almost completely diminish even if the temperature satisfy $T/Q\sim 10^{-13}$\,K. 

The outgoing optical phase spectrum (normalised by the spectrum without gravity effect at $\omega_m$) under the different prescriptions is depicted in Fig.\ref{fig:spectrum}. The upper panel shows the spectrum around $\omega_m$ and the signal peak appears at $\omega_m$ under the causal-conditional prescription, which corresponds to the resonant oscillation frequency of the $\langle\hat{x}\rangle$. The post-selection prescription also has a featured signal near $\omega_m$. The spectrum around $\omega_q$ is shown in the middle panel, where there is no featured signal for the causal-conditional prescription. Moreover, taking into account the thermal noise will make it very difficult to distinguish the spectrum for quantum gravity and semi-classical gravity under causal conditional prescription.

\subsubsection{Pondermotive squeezing}
Another phenomenon in this optomechanical system under the causal-condition prescription is the pondermotive squeezing\,\cite{Corbitt2006}. Pondermotive squeezing is the radiation-pressure induced correlation between the phase and amplitude quadratures of the outgoing optical field, which was experimentally demonstrated in\,\cite{Aggarwal2020,Cripe2019,Purdy2017,Brooks2012,Militaru2022}. When the self-gravity is quantum, to the leading order, the Hamiltonian has the same form as that when there is no gravity effect, so does the pondermotive squeezing spectrum.

\begin{figure}[h]
\centering
\includegraphics[width=0.49\textwidth]{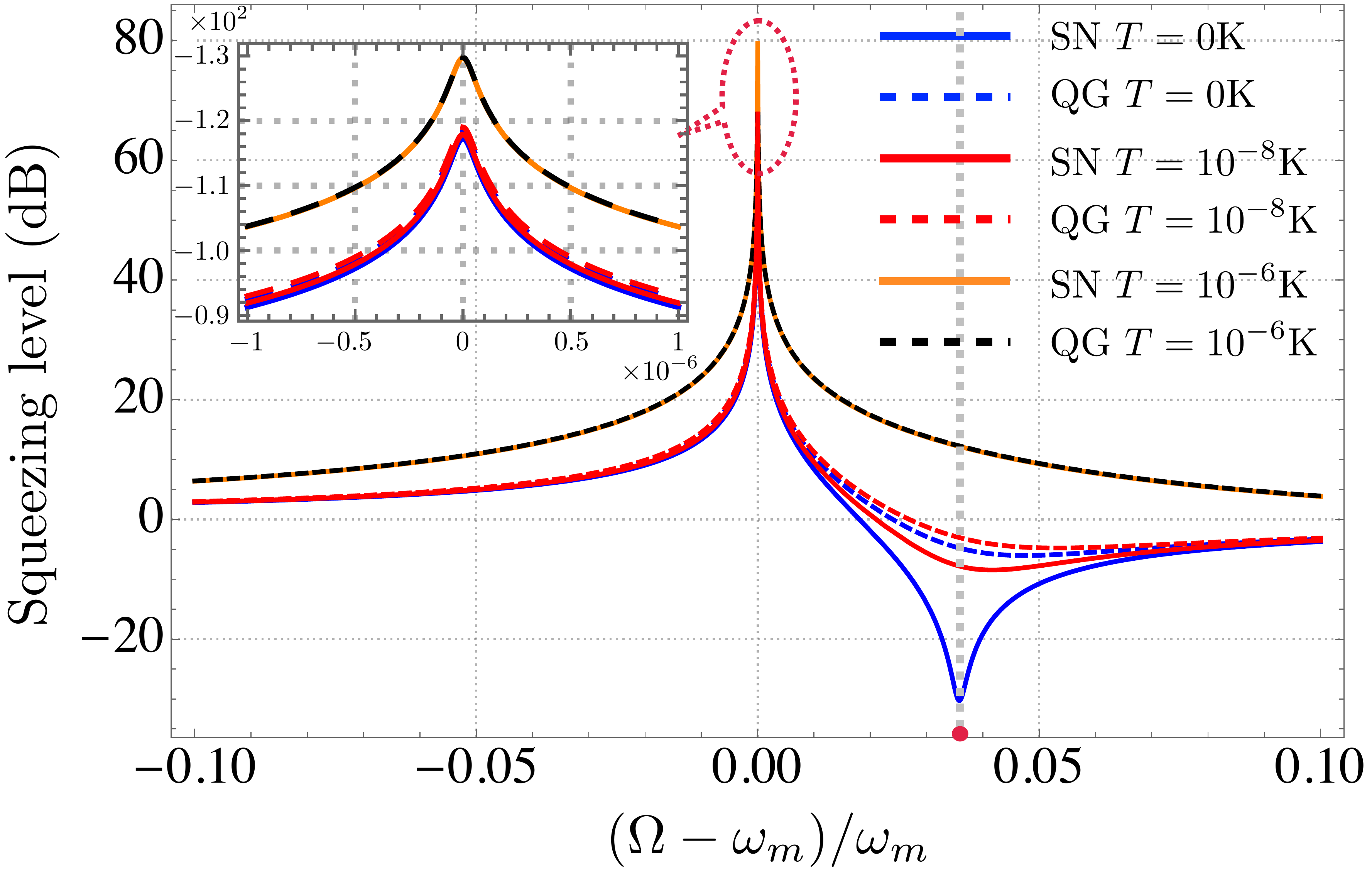}
\caption{Pondermotive squeezing spectrum for the optomechanical system with a single mirror exerted by its self-gravity. Solid curves: Pondermotive squeezing in the semi-classical SN theory, under the causal-conditional prescription.  Dashed curves: pondermotive squeezing when the gravity is quantum. The red spot denotes the frequency where there will be a distinctive feature. Thermal effect is also considered which shows that the semi-classical gravity and quantum gravity can only be distinguished when the parameters setting is very extreme.}\label{fig:pondermotive_squeezing_self}
\end{figure}

However, when the self-gravity is semi-classical with Hamiltonian described by Eq.\,\eqref{eq:Hamiltonian_self_gravity}, there will be a different pondermotive squeezing spectrum under the causal-conditional prescription (see Fig.\,\ref{fig:pondermotive_squeezing_self}). This difference happens when the self-gravity effect is strong $\omega_{\rm SN}\geq \omega_m$, which will diminish in case the gravity effect is weak, e.g. in the mutual gravity case shown later. %see Fig.\,\eqref{fig:pondermotive_squeezing_self}. 

%\R{Some comment on the result: the self gravity situation does not have a good feature?? how to speak out the pondermotive squeezing feature?? Some formulae?}

We perform the following steps to calculate the pondermotive squeezing effect under the causal-conditional prescription. The output optical field has a spectrum given as $S_{a_\theta a_\theta}=S_{a_1a_1}\cos^2\theta+S_{a_2a_2}\sin^2\theta+2{\rm Re}[S_{a_1a_2}]\sin\theta\cos\theta$ . The $\hat a_\theta$ is the $\theta$-quadrature of the output field, in which $\hat a_\theta=\hat a_1\cos\theta+\hat a_2\sin\theta$ with $\hat a_{1/2}$ represents the amplitude/phase quadrature of the output field, and the $\theta$ is the homodyne angle. %The readout values of $\hat a_{\theta}$ relate to the conditional motion via:
%\be
%\tilde{a}_{\theta}=dW_\theta/\sqrt{2}dt+\alpha\langle \hat x\rangle_c\sin\theta, 
%\ee
%where $dW_\theta=dW_1\cos\theta+dW_2\sin\theta$ and also satisfies Ito calculus $dW_\theta^2=dt$.
Optimisation to the $\theta$ leads to the quadrature with minimum uncertainty (i.e. the pondermotive squeezing) shown in Fig.\,\ref{fig:pondermotive_squeezing_self}. Taking into account the thermal noise effect, this result shows that the quantum gravity and semi-classical gravity can be distinguished using the pondermotive squeezing spectrum for the optomechanical system with mirror exerted by its self-gravity, only when the environmental temperature takes an extremely low value.

\section{Optomechanical system with two mirrors interacting via mutual gravity}\label{sec.4}
\subsection{General Discussions}
Two mirrors can be coupled via mutual gravitational force. In the discussion below, we ignore the self-gravity effect and only consider the mutual gravity for illustrative purpose. In the SN theory, the corresponding Hamiltonian can be written as\,\cite{Miao2020}:
\be\label{eq:Hamiltonian_mutual}
\begin{split}
\hat H_{\rm SN}=\sum_{A/B}\left[\frac{\hat p^2_{A/B}}{2m}+\frac{m\omega_m^2}{2}\hat x^2_{A/B}-\mathcal{C}(\hat x_{A/B}-\langle\hat x_{B/A}\rangle_c)^2\right].
\end{split}
\ee
Here, we have $\mathcal{C}=m\tilde\Lambda G\rho$ with $\tilde\Lambda$ decided by the geometric shape of the mirror, and $\rho$ is the mirror matter density.

Under the causal-conditional prescription, the quantum expectation value here is for the conditional quantum state of the mirrors. Since different measurement schemes of the optical fields will lead to different conditional mechanical quantum states, the configuration topology of the optomechanical device is important, as we will show later.  In this work, two different configurations will be studied: (1)\,\emph{Folded interferometer configruation} where the projective measurement is performed on the common and differential optical fields\,\cite{Datta_2021} (see Fig.\,\ref{fig:interferometer_setup}). As we shall see, the folded interferometer configuration can be mapped onto a single-mirror problem. (2)\,\emph{Linear cavity configuration} where the projective measurement is directly performed on the outgoing optical field reflected by each cavity\,\cite{Miao2020} (Fig.\,\ref{fig:linear_cavity_setup}), which can not be mapped onto a single-mirror problem and the quantum measurement induced correlation mentioned in the \emph{Introduction} will appear.  In both two cases, the interaction Hamiltonian between the optical fields and the mechanical motion takes the same form
\be
\hat H_{\rm int}=-\hbar\alpha(\hat a_1\hat x_A+\hat b_1\hat x_B),
\ee
where we assumed that the linear optomechanical coupling strengths $\alpha$ of both mirrors are the same. We also assume that both cavities have no detuning to the frequency of their pumping lasers, which means there is no dynamical back-action in the optomechanical system.

Another important point that needs to be discussed is the treatment of environmental noises such as thermal noise.  It is known that all environmental noises are \emph{fundamentally speaking} quantum mechanical since these environmental degrees of freedom are \emph{quantum mechanical}.  %On one hand, when the environmental temperature is low for a high-$Q$ mechanical oscillator so that the $\bar n_{\rm th}/Q\sim k_BT_{\rm env}/\hbar\omega_m Q_m<1$, the environmental effect behaves quantum mechanically. 
Since the classical gravity can not convey quantum mechanical information,  the quantum environmental effect will only apply to the fluctuation of the A/B mirror on its own thereby does not contribute to the mutual correlation between the output fields shown in Fig.\,\ref{fig:linear_cavity_setup}. Mathematically, there will be additional thermal terms in the evolution equation for the second-order correlation functions (see the Appendix), while the evolution for the first-order moments is unchanged. Note that in this case, the mutual correlation of two mirrors satisfies $V^c_{(x/p)_A(x/p)_B}=0$, which  %On the other hand, if the $T_{\rm env}/Q_m$ is high so that $\bar n_{\rm th}\gg 1$, the thermal environment can be treated classically so that phenomenological Langevin terms can be added to the evolution of \emph{the first-order moments}, while the equation for the second-order moments will remain to be the same (which can be derived from the stochastic master equation). The treatment of thermal noise in this work will follow the above-mentioned methods.
%The above treatment brings us an additional bonus which can be seen clearer later. On one hand, in the low $T_{\rm env}/Q_m$ limit, the classical gravity can not establish the correlations between the two quantum mirrors, therefore $V^c_{(x/p)_A(x/p)_B}=0$. On the other hand, in the high $T_{\rm env}/Q_m$ limit, the thermal environmental effect contributes as a classical Langevin force and does not affect the second-order conditional quantum moments, which means $V^c_{(x/p)_A(x/p)_B}=0$. 
%This argument can also be applied to mirror state preparation process, where the initial states of the two mirrors do not have mutual quantum correlations in both two cases. 
means that the Ricatti equation that describes the evolution of the second-order moments can be treated separately. 

\begin{table}[h!]
    \centering
    \begin{tabular}{|c|c|c|}
    \hline
Parameters&Symbol&Value\\
\hline
Mirror mass&$M$&$10^{-3}$\,kg\\
\hline
Mirror bare frequency&$\omega_m/(2\pi)$&$0.5\,{\rm Hz}$\\
\hline
SN frequency&$\omega_{g}/(2\pi)$&$2\times10^{-4}\,{\rm Hz}$\\
\hline
Quality factor&$Q_m$&$3\times10^7$\\
\hline
Mechanical damping&$\gamma_m/(2\pi)$&$1.67\times10^{-8}\,{\rm Hz}$\\
\hline
Environmental temperature &$T$&$300\,{\rm K}$\\
\hline
Optical wavelength&$\lambda$&$1064\,{\rm nm}$\\
\hline
Cavity Finesse&$\mathcal{F}$&$4000$\\
\hline
Intra-cavity power&$P_{\rm cav}$&$2000\,{\rm W}$\\
\hline
    \end{tabular}
    \caption{The parameters of the optomechanical system with two mirrors interacting via mutual SN gravity.}
    \label{tab:mutual_gravity}
\end{table}

\subsection{Folded interferometer configuration}
The conditional mechanical quantum state in the SN theory can also be prepared by projecting the optical field in the common and differential mode basis. For example, in a folded interferometer configuration, the optical field being measured is: $\hat c=1/\sqrt{2}(\hat a+\hat b)$ and $\hat d=1/\sqrt{2}(\hat a-\hat b)$. The common and differential motional degrees of freedom are: $\hat x_\pm=1/\sqrt{2}(\hat x_A\pm\hat x_B)$ and $\hat p_\pm=1/\sqrt{2}(\hat p_A\pm\hat p_B)$. Written in terms of these common and differential optical/mechanical modes, the Hamiltonian Eq.\,\eqref{eq:Hamiltonian_mutual}
can be transformed into:
\be
\begin{split}
\hat H_{\pm}=\frac{\hat p^2_{\pm}}{2m}+\left(\frac{m\omega_m^2}{2}-\mathcal{C}\right)\hat x^2_{\pm}\pm 2\mathcal{C}\hat x_{\pm}\langle \hat x_{\pm}\rangle-\hbar\alpha \hat c_1/\hat d_1\hat x_\pm,
\end{split}
\ee
where $\mathcal{C}$ is the mutual-gravity Schr{\"o}dinger-Newton coefficient, $\hat c_1/\hat d_1$ means $\hat c_1$ or $\hat d_1$.
This Hamiltonian shows that the common and differential modes are completely decoupled from each other, which means we can treat them independently.

Using the Stochastic master equation, the conditional expectation value can be obtained as:
\be\label{eq:Bexpectation}
\begin{split}
&d\langle \hat x_+\rangle=\frac{\langle \hat p_+\rangle}{m}dt+\sqrt{2}\alpha V^c_{x_+x_+}dW_c+\sqrt{2}\alpha V^c_{x_+x_-}dW_d,\\
&d\langle \hat p_+\rangle=-m\omega_m^2\langle x_+\rangle dt+2\mathcal{C}(\langle \hat x_+\rangle_c-\langle \psi_m|\hat x_+|\psi_m\rangle)dt\\
&\qquad\qquad+\sqrt{2}\alpha V^c_{x_+p_+}dW_c+\sqrt{2}\alpha V^c_{p_+x_-}dW_d,\\
&d\langle \hat x_-\rangle=\frac{\langle \hat p_-\rangle}{m}dt+\sqrt{2}\alpha V^c_{x_-x_-}dW_d+\sqrt{2}\alpha V^c_{x_-x_+}dW_c,\\
&d\langle \hat p_-\rangle=-m\omega_m^2\langle \hat x_-\rangle dt+2\mathcal{C}(\langle \hat x_-\rangle_c+\langle \psi_m|\hat x_-|\psi_m\rangle)dt\\
&\qquad\qquad+\sqrt{2}\alpha V^c_{x_-p_-}dW_d+\sqrt{2}\alpha V^c_{x_-p_+}dW_c,
\end{split}
\ee
where $dW_{c/d}$ are the Ito-terms for measuring the optical common/differential fields. The form of $\langle \psi_m|\hat x_\pm|\psi_m\rangle$ depends on the measurement prescription. Under pre and post-selection prescriptions, we have $\langle \psi_m(t_{i})|\hat U_{|\psi_m(t_{i})\rangle}^\dag(t,t_i)\hat x_\pm\hat U_{|\psi_m(t_{i})\rangle}(t,t_i)|\psi_m(t_{i})\rangle$ and $\langle \psi_m(t_{f})|\hat U_{|\psi_m(t_{f})\rangle}(t,t_i)\hat x_\pm\hat U^\dag_{|\psi_m(t_{f})\rangle}(t,t_i)|\psi_m(t_{f})\rangle$, respectively. However, under the causal-conditional prescription, we have $\langle \psi_m|\hat x_\pm|\psi_m\rangle=\langle \hat x_{\pm}\rangle_c$. Therefore the mutual gravity terms in Eq.\,\eqref{eq:Bexpectation} can be simplified to be zero and $4\mathcal{C}\langle \hat x_{-}\rangle_c$ for $\langle \hat p_+\rangle$ and $\langle \hat p_-\rangle$, respectively. This means that the conditional mean of the common and differential modes will evolve with frequency $\omega_m$ and $\omega_-=\sqrt{\omega^2_m-4\mathcal{C}/m}$.

\begin{figure}[h]
\centering
\includegraphics[width=0.4\textwidth]{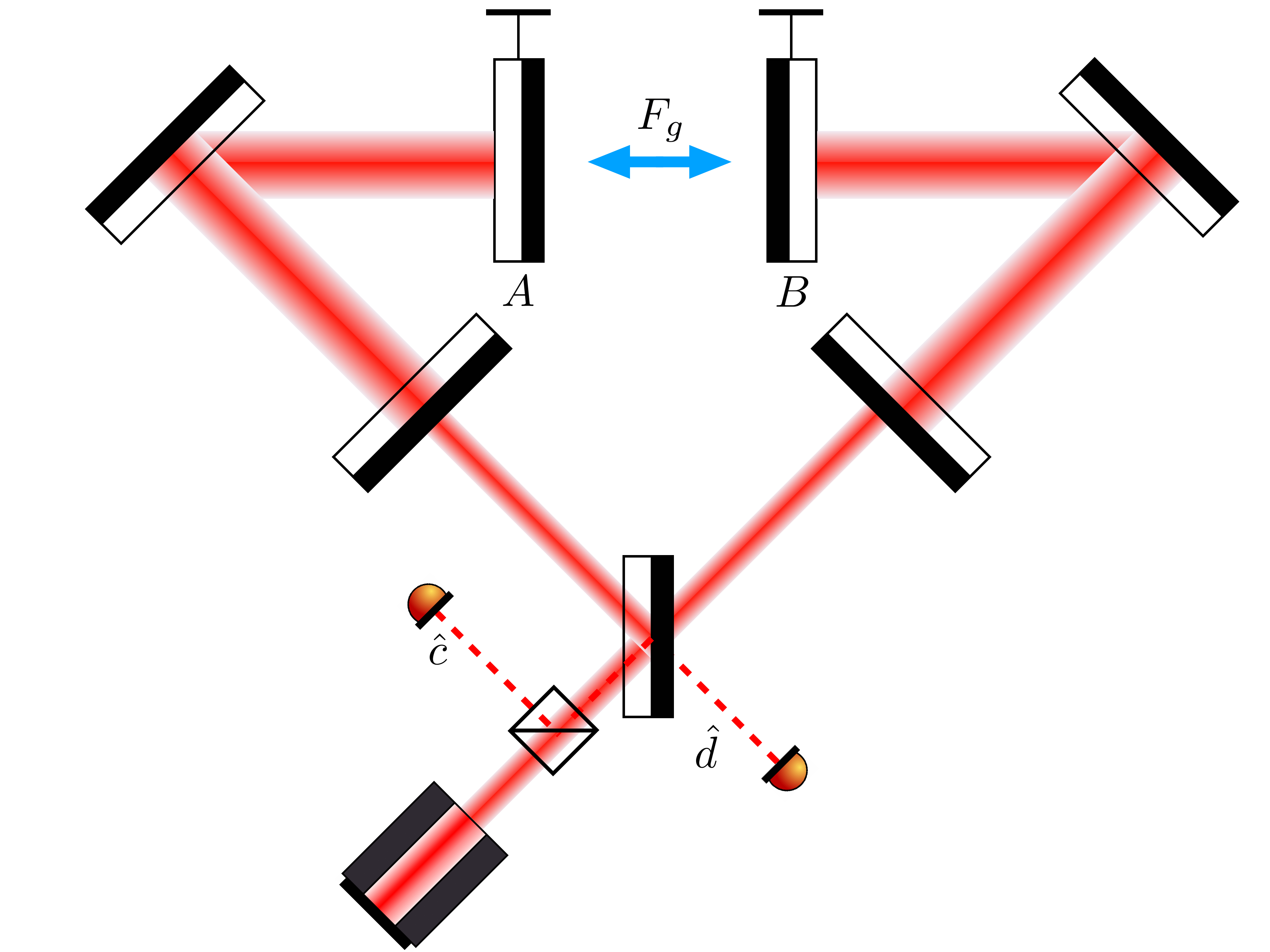}\\
\includegraphics[width=0.5\textwidth]{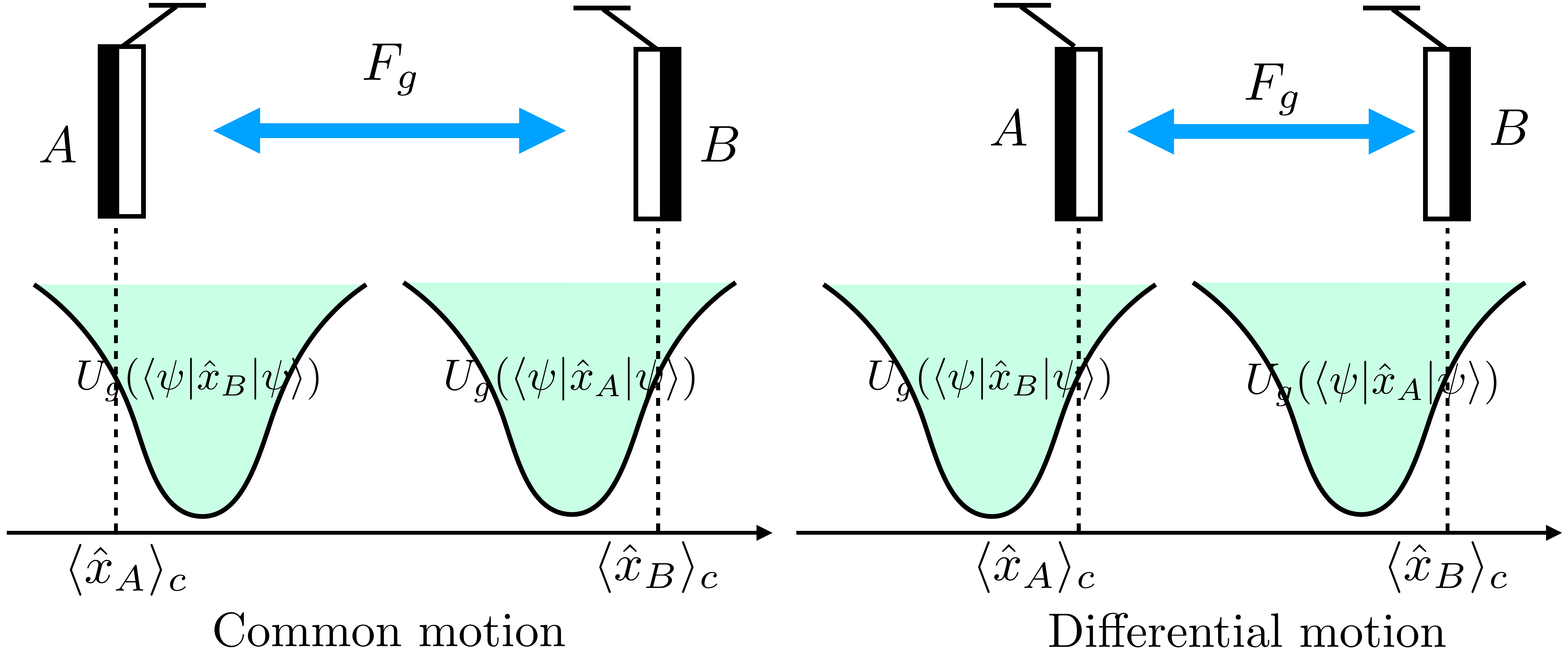}
\caption{Upper panel: Folded interferometer configuration. The optical field is projected into the common mode $\hat c$ and differential mode $\hat d$. The common and differential motions of these two mirrors are correspondingly prepared into the quantum states, of which the evolution can be separated. Lower panel: the gravitational potential felt by the quantum trajectories of common and differential motions under the pre/post selection prescription. }\label{fig:interferometer_setup}
\end{figure}

The above equations for the quantum trajectory are physically transparent, see the lower panel of Fig.\,\ref{fig:interferometer_setup}. Under the pre/post-selection prescription, the random quantum trajectory of mirror A/B with the mean displacement $\langle \hat x_{A/B}\rangle_c$ will feel a harmonic-like potential $U_g(\langle\psi_m|\hat x_{A/B}(t)\psi_m\rangle)$ generated by the classical gravitational force of mirror B/A at pre/post-selection state and their suspension systems, which follows a deterministic evolution. From Fig.\,\ref{fig:interferometer_setup}, it is clear that the potential force felt by common/differential motion trajectory will drive common/differential motion trajectory themselves, which contributes to the $\mathcal{C}$-dependence term in Eq.\,\eqref{eq:Bexpectation}. These terms will change the resonant frequency of the two motional modes. However, under the causal-conditional prescription, the potential also follows a quantum trajectory. Therefore there is no additional restoring force contributed by the mutual gravity for the common motion, which is different from the case of the differential motion. 

This leads to a different result compared to that predicted in\,\cite{Datta_2021}, which focuses on the quantum gravity and the semi-classical gravity is only briefly discussed without specifying the detailed quantum measurement processes. In\,\cite{Datta_2021}, the features of quantum gravity manifest in the pondermotive squeezing of the outgoing field. They show that there is a frequency shift between the pondermotive squeezing spectrum for the common and differential output fields in the quantum gravity theory, while there is no such frequency shift for the semi-classical gravity.  In the following, a similar calculation will be performed for both quantum gravity and the Schroedinger-Newton theory under the causal-conditional prescription.
%Under the causal conditional prescription, the common mode in the SN theory has the same behaviour as that in the quantum gravity which has been studied before in Ref.[XXX].

%\R{The output optical field (take the differential $\hat d$-field as an example) has a spectrum given as $S_{d_\theta d_\theta}=S_{d_1d_1}\cos^2\theta+S_{d_2d_2}\sin^2\theta+2S_{d_1d_2}\sin\theta\cos\theta$ . The $\hat d_\theta$ is the $\theta$-quadrature of the output field, in which $\hat d_\theta=\hat d_1\cos\theta+\hat d_2\sin\theta$ with $\hat d_{1/2}$ represents the amplitude/phase quadrature of the output differential mode, and the $\theta$ is the homodyne angle. The readout values of $\hat d_{\theta}$ relate to the conditional motion via:
%\be
%\tilde{d}_{\theta}=dW_\theta/\sqrt{2}dt+\alpha\langle \hat q_-\rangle_c\sin\theta, 
%\ee
%where $dW_\theta=dW_1\cos\theta+dW_2\sin\theta$ and also satisfy Ito calculus $dW_\theta^2=dt$. 

\begin{figure}[h!]
\centering
\includegraphics[width=0.45\textwidth]{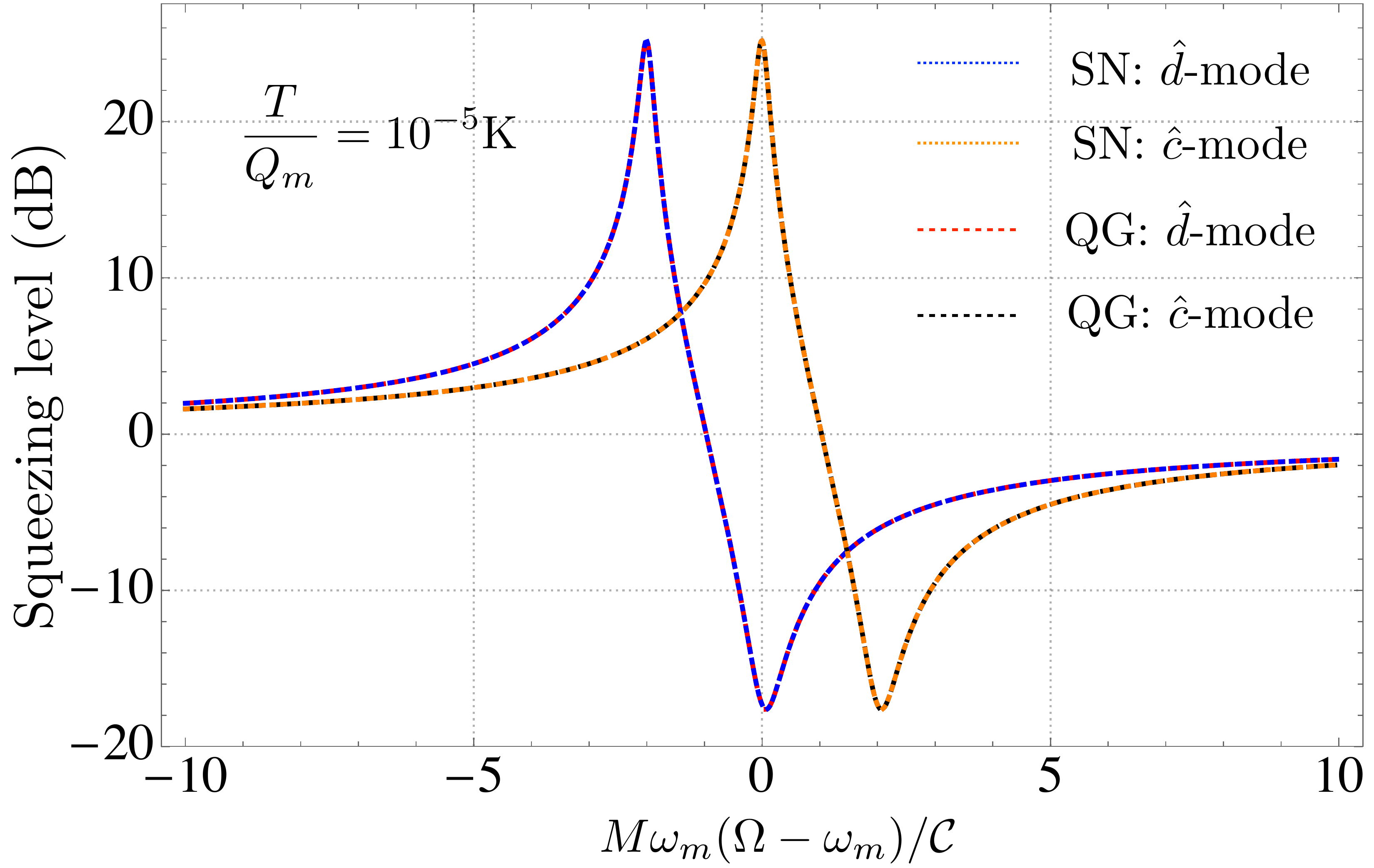}
\includegraphics[width=0.45\textwidth]{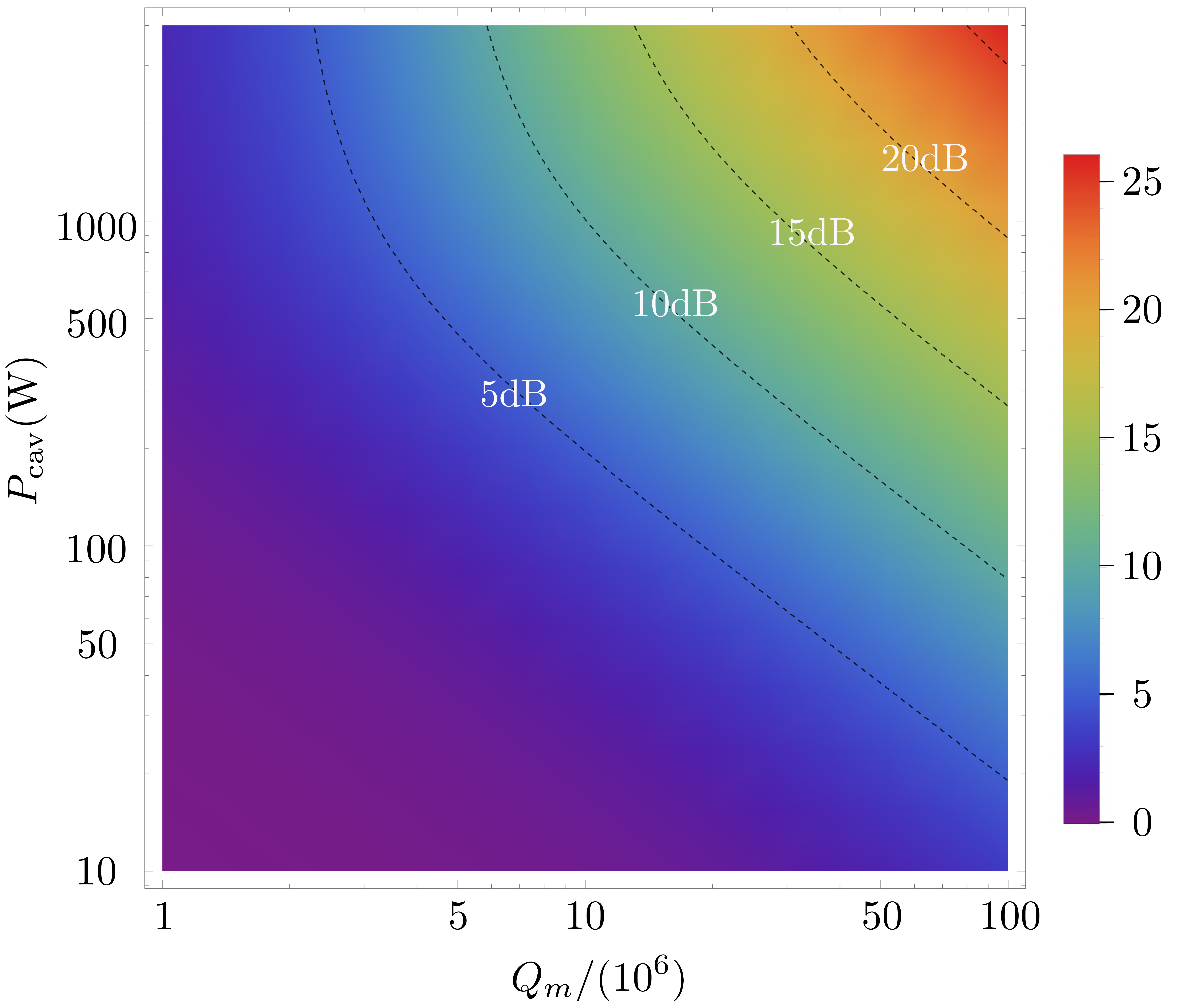}
\includegraphics[width=0.45\textwidth]{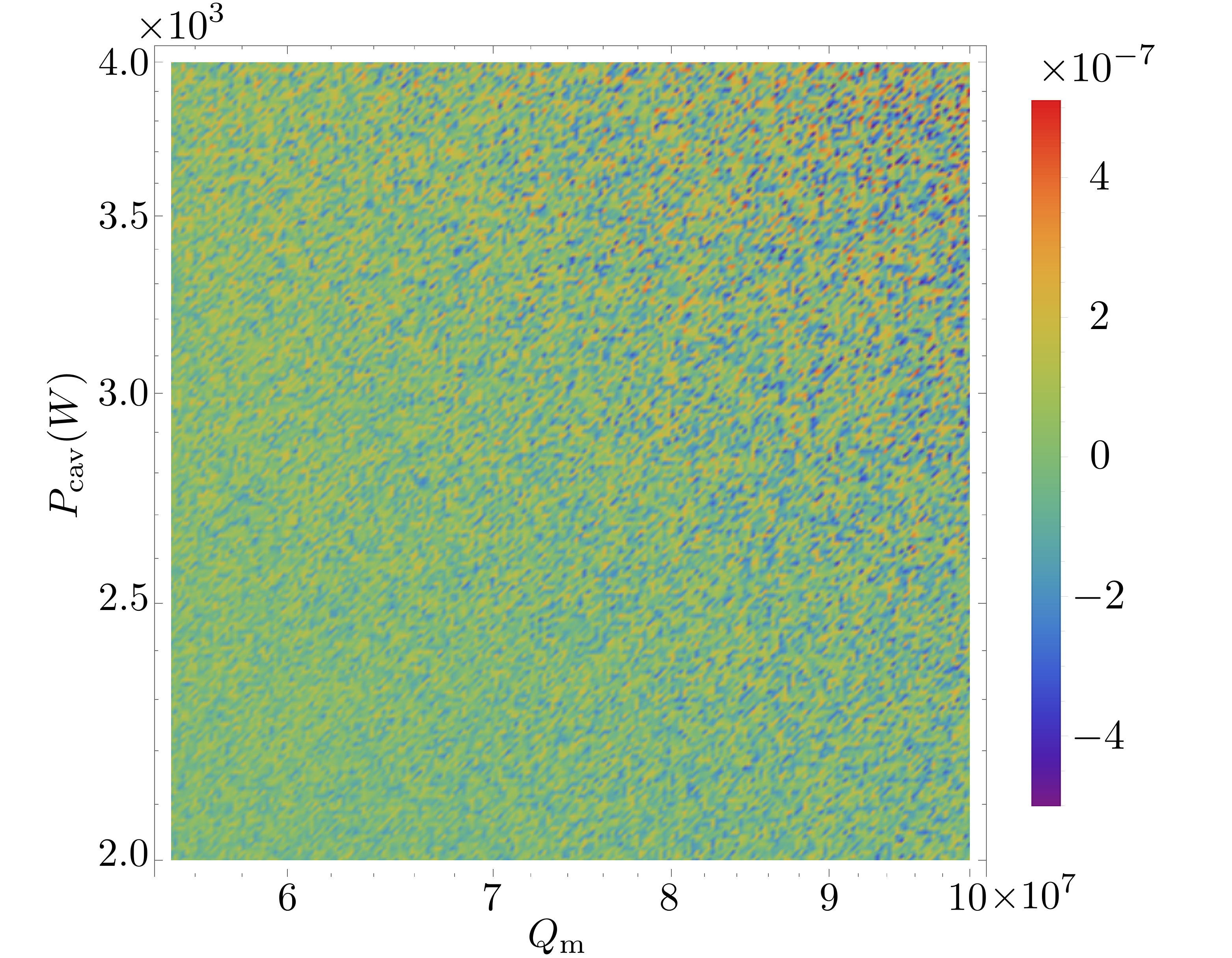}
\caption{Upper pannel: The squeezing spectrum of the outgoing field from common and differential modes, including both the quantum gravity case and the SN case, calculated using the casual-conditional prescription. Middle pannel: the dependence of maximum squeezing level on intra-cavity power $P_{\rm cav}$ and the mechanical quality factor $Q_m$. Lower panel: The (almost zero) difference between pondermotive squeezing level $\Delta S^{\rm SN-QG}_{\rm sq}$ in SN theory and that in quantum gravity, where we focus on the peak value at $\Omega=\omega_m$. We only choose a sub-region of parameter space in the middle panel with the difference is around $10^{-7}$\,dB, while the value in other region of parameter space is much smaller.}
\label{fig:pondermotive_squeezing}
\end{figure}

Following the method in\,\cite{Datta_2021}, we can measure the pondermotive squeezing effect in this mutual-gravity optomechanical system. For the common/differential output channel, optimisation to the homodyne angle $\theta$ leads to the quadrature with minimum uncertainty (i.e. the pondermotive squeezing) shown in Fig.\,\ref{fig:pondermotive_squeezing}\,\footnote{Note that the work in\,\cite{Datta_2021} only present the squeezing level at $\omega_m$ for the quantum gravity case without presenting the full squeezing spectrum.}, where we show the minimum quadrature uncertainty (in terms of the squeezing level) of the output $\hat c$ and $\hat d$ states when $T/Q_m=10^{-5}{\rm K}$ under the causal-conditional prescription.  As shown clearly in Fig.\,\ref{fig:pondermotive_squeezing}, one can not distinguish semi-classical gravity from quantum gravity by investigating the pondermotive squeezing spectrum under the causal-conditional prescription, that is, the frequency shift still exists for the semi-classical gravity. This result can be understood from the similarity between Eq.\,\eqref{eq:Bexpectation} and the Heisenberg equation of motion for the mirrors in the quantum gravity theory\,\cite{Datta_2021}. In addition, there is a very tiny (almost indistinguishable) difference on the numerical value of the squeezing level for the semi-classical gravity and quantum gravity, which originates from the difference of steady solution of second-order conditional moments in these two cases (for example, the $\omega_q$ in Eq.\,\eqref{eq:steadyvariance} should be replaced by $( \sqrt{\omega_m^2-2\mathcal{C}/m},\omega_-)$ for the differential mode in the SN theory and quantum gravity, respectively).

For completeness, we present the calculation details and the results for the semi-classical SN theory under the pre/post selection prescription in the Supplementary Material of this paper, which would certainly have a different pondermotive squeezing spectrum compared to that under the causal-conditional prescription. Moreover, only the spectrum under the pre-selection prescription fits the results in\,\cite{Datta_2021}, where there is no frequency shift between the $\hat c$ and $\hat d$-spectrum.

In summary, in contrast to the conclusion made in\,\cite{Datta_2021}, the pondermotive squeezing is not an ideal figure of merit for testing the semi-classical gravity theory under the conditional prescription and the quantum gravity.

\subsection{Linear cavities configuration} 
\begin{figure}[h]
\centering
\includegraphics[width=0.5\textwidth]{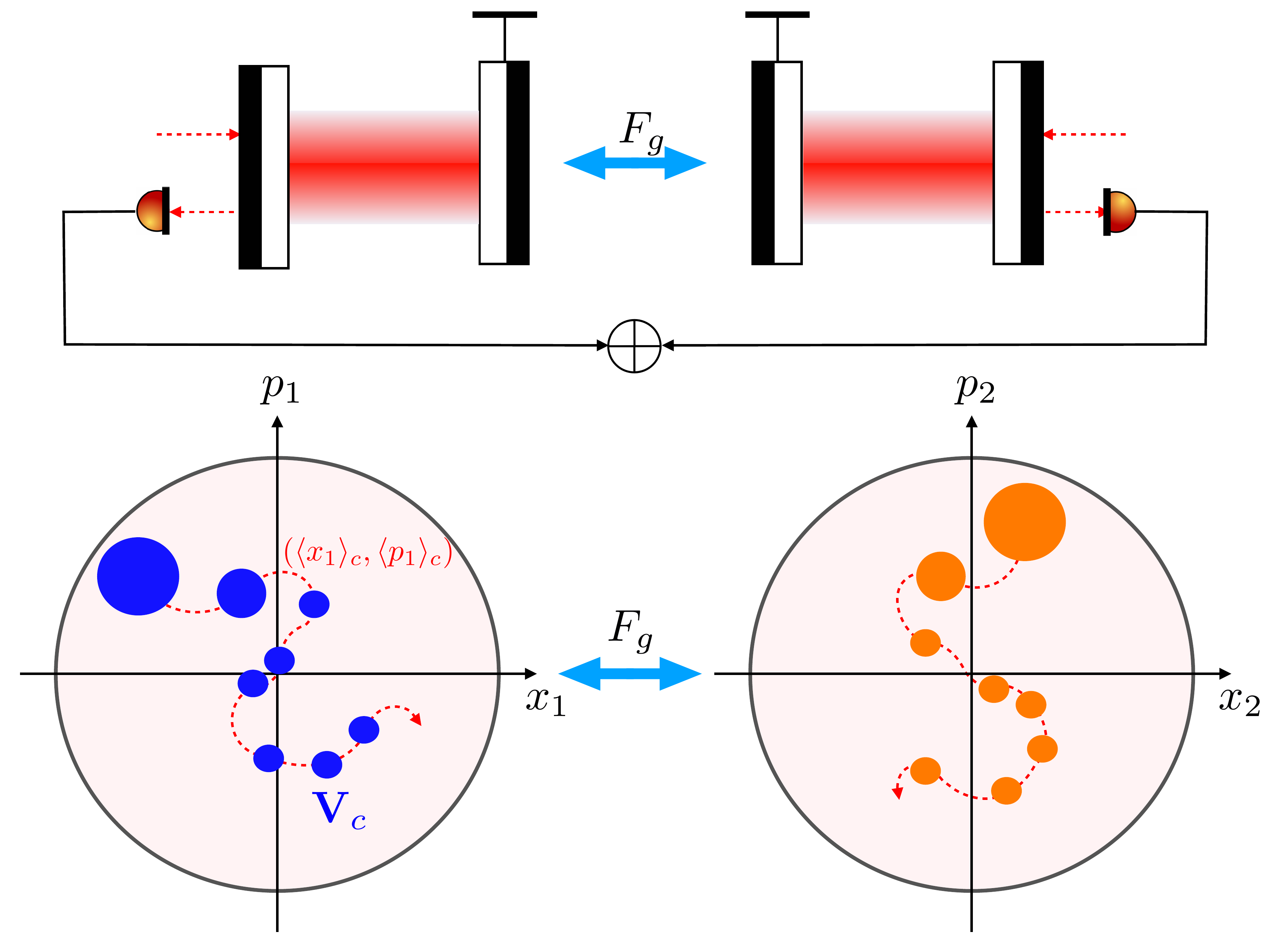}
\caption{Correlations induced by quantum measurement. The projection of the joint optomechanical state onto the optical Hilbert space can produce a mechanical quantum trajectory, which can source the semi-classical gravity and affect the motion of the other mirror. The correlation between the mechanical motion of the two mirrors will manifest itself in the correlation of the outgoing fields.}\label{fig:linear_cavity_setup}
\end{figure}

Now we switch to the linear cavities configuration shown in Fig.\,\ref{fig:linear_cavity_setup}, whose main physical scenario can be summarised as follows. If the gravity is quantum and thereby can mediate quantum information, there will be a \emph{quantum correlation} between the two output light fluctuations, which has been studied in\,\cite{Miao2020}. In contrast, the work in\,\cite{Miao2020} argues that if the gravity is classical and sourced by the quantum expectation value of the energy-momentum of the mirror A, then the mirror B will not feel the fluctuating gravity force sourced by A. This is because the mean value of the first mirror position is zero. Finally, the work in\,\cite{Miao2020} concludes that there will be no correlation between the two output light noises.

However, although there is no quantum correlation between the two output light fields, there will be \emph{quantum measurement induced classical correlations} under the causal-conditional prescription.   Continuous monitoring of the mechanical degree of freedom via projective measurement on the optical fields collapses the joint optomechanical quantum state and thereby preparing the so-called \emph{conditional quantum state} of the mirror motion denoted as $|\psi_c\rangle$. The mean value of the mirror displacement $\bar x_c=\langle\psi_c|\hat x|\psi_c\rangle$ and momentum $\bar p_c=\langle\psi_c|\hat p|\psi_c\rangle$ will follow a stochastic \emph{quantum trajectory}\,\cite{Rossi2019}, which is determined by the measurement records (see Fig.\,\ref{fig:linear_cavity_setup}). Therefore even if the gravity is classical (i.e., the mirror motion is described by SN theory), the ``stochastic motion" of the $(\bar x_c,\bar p_c)$ of mirror A in the phase space can still source a gravitational force acting on the mirror B (vice versa), which then can be further recorded by the light field monitoring the position of mirror B. In this way, the two light fields establish a \emph{classical correlation} via the \emph{classical gravitational interaction sourced by the quantum trajectory induced by the wave function collapse}, which will be quantitatively studied as follows.

Suppose we measure the arbitrary quadratures of the outgoing optical fields:
\be
\begin{split}
\hat{a}_{\theta A}=\hat{a}_{1A}\cos\theta_A+\hat{a}_{2A}\sin\theta_A,\\
\hat{a}_{\theta B}=\hat{a}_{1B}\cos\theta_B+\hat{a}_{2B}\sin\theta_B,
\end{split}
\ee
then the quantum trajectories satisfy the following equations:
%\begin{equation}\label{eq:two_quantum_trajectory}
%\begin{split}
% &d\langle\hat{x}_A\rangle=\frac{\langle \hat{p}_A\rangle}{M}dt+\sqrt{2}\alpha V_{x_Ax_A}dW_A+\sqrt{2}\alpha V_{x_Ax_B}dW_B\\
% &d\langle\hat{p}_A\rangle=-M\omega_q^2\langle\hat{x}_A\rangle dt-\gamma_m\langle\hat{p}_A\rangle dt+M\omega_g^2\langle\hat{x}_B\rangle dt+\sqrt{2}\alpha V_{x_Ap_A}dW_A+\sqrt{2}\alpha V_{x_Bp_A}dW_B\\
% &d\langle\hat{x}_B\rangle=\frac{\langle \hat{p}_B\rangle}{M}dt+\sqrt{2}\alpha V_{x_Ax_B}dW_A+\sqrt{2}\alpha V_{x_Bx_B}dW_B\\
% &d\langle\hat{p}_B\rangle=-M\omega_{q}^2\langle\hat{x}_B\rangle dt-\gamma_m\langle\hat{p}_B\rangle dt+M\omega_g^2\langle\hat{x}_A\rangle dt+\sqrt{2}\alpha V_{x_Ap_B}dW_A+\sqrt{2}\alpha V_{x_Bp_B}dW_B
%\end{split}
%\end{equation}
\be\label{eq:linear_cavity_trajectory}
d\mathbf{x}_{A/B}=\mathbb{G}_m^q\mathbf{x}_{A/B}dt+\mathbb{G}_g\mathbf{x}_{B/A}dt+\sqrt{2}\alpha\mathbb{V}_{A/B}d\mathbf{W},
\ee
where $\mathbf{x}_{A/B}=[\langle \hat x_{A/B}\rangle_c,\langle \hat p_{A/B}\rangle_c]^T$, $d\mathbf{W}=[dW_A,dW_B]^T$. The $\mathbb{G}_m^q$ is the mechanical response under the Schroedinger-Newton theory:
\be\label{eq:G-matrix}
\mathbb{G}_m^q=\left[
\begin{array}{cc}
0&1/M\\
-M\omega_q^2&-\gamma_m
\end{array}
\right],
\quad
\mathbb{G}_g=\left[
\begin{array}{cc}
0&0\\
M\omega_g^2&0
\end{array}
\right],
\ee
and
\be\label{eq:V-matrix}
\begin{split}
&\mathbb{V}_{A}=\left[
\begin{array}{cc}
V_{x_Ax_{A}}\sin\theta_A&V_{x_{A}x_B}\sin\theta_B\\
V_{x_Ap_{A}}\sin\theta_A+(\hbar/2)\cos\theta_A&V_{x_Bp_{A}}\sin\theta_B
\end{array}
\right],\\
&\mathbb{V}_{B}=\left[
\begin{array}{cc}
V_{x_Ax_{B}}\sin\theta_A&V_{x_{B}x_B}\sin\theta_B\\
V_{x_Ap_{B}}\sin\theta_A&V_{x_Bp_{B}}\sin\theta_B+(\hbar/2)\cos\theta_B
\end{array}
\right].
\end{split}
\ee
In the mutual gravity case, we redefine $\omega_q=\sqrt{\omega_m^2-\omega_g^2}$, where $\omega_g=\sqrt{2\mathcal{C}/M}$. The $\gamma_m$ term here represents the mechanical loss due to the coupling with the thermal bath, where the corresponding thermal force term is discussed in detail in the Supplementary Material.

The second order correlations for the A and B mirrors are\,\footnote{The lengthy concrete form will be shown in the appendix.}:
\be
\begin{split}
\mathcal{\hat D}\mathbf{V}_A=-2\alpha^2[V_{x_Ax_B}^2,V_{x_Ax_B}V_{p_Ax_B},V_{x_Bp_A}^2]^T\\
\mathcal{\hat D}\mathbf{V}_B=-2\alpha^2[V_{x_Ax_B}^2,V_{x_Ax_B}V_{p_Ax_B},V_{x_Bp_A}^2]^T
\end{split}
\ee
where $\mathbf{V}_{A}=[V_{x_Ax_A},V_{x_Ap_A},V_{p_Ap_A}]^T$ and the $\mathcal{\hat D}\mathbf{V}_A$ represents the same equation as Eq.\,\eqref{eq:steadyvariance} (similar for the $\mathbf{V}_{B}$) for arbitrary measurement angle $\theta_A,\theta_B$. The right-hand side is the contribution of the mutual correlations between these two mirrors. As we shall show in the Supplementary Material, the evolution equations for these mutual correlations do not depend on the elements of $\mathbf{V}_{A/B}$, which is different 
from the quantum gravity case. Therefore, in the ideal case, if these two mirrors are \emph{prepared in the uncorrelated initial state}, then the mutual correlations are always zero. In this case, the steady value of the second-order correlation has the same form as the one given in Eq.\,\eqref{eq:steadyvariance}.

The mutual correlation induced by the classical mutual gravity between the two mirrors' quantum trajectories now actually comes from the $d\langle p\rangle_{A/B}\supset M\omega_g^2\langle x\rangle_{B/A}dt$ (i.e., the $\mathbb{G}_g$ term in the Eq.\,\eqref{eq:linear_cavity_trajectory}), where the $\langle \hat x\rangle_{B/A}$ has randomness induced by the quantum measurement of light field. For example, if we measure the phase qudrature of the two outgoing light fields:
\be\label{eq:a2AB}
\tilde{a}_{2A/B}=\alpha\langle x_{A/B}\rangle_c+dW_{A/B}/\sqrt{2}dt,
\ee
their mutual correlation is only contributed by the classical gravitational interaction between the two quantum trajectories. More explicitly, the conditional mean displacement can be formally solved as:
\be\label{eq:xAexpectation}
\begin{split}
&\langle \hat x_{A}(t)\rangle_c=e^{-\gamma_mt/2}x^{(0)}_{A}(t)+\\
&\sqrt{2}\alpha\int^t_0dW_A(s)\left[V_{x_Ax_A}\cos\omega_m(t-s)+V_{x_Ap_A}\frac{\sin\omega_m(t-s)}{M\omega_m}\right]+\\
&\sqrt{2}\alpha\int^t_0dW_A(s)\left[V_{x_Ax_A}\cos\omega_-(t-s)+V_{x_Ap_A}\frac{\sin\omega_-(t-s)}{M\omega_-}\right]+\\
&\sqrt{2}\alpha\int^t_0dW_B(s)\left[V_{x_Bx_B}\cos\omega_m(t-s)-V_{x_Bp_B}\frac{\sin\omega_m(t-s)}{M\omega_m}\right]+\\
&\sqrt{2}\alpha\int^t_0dW_B(s)\left[V_{x_Bx_B}\cos\omega_-(t-s)-V_{x_Bp_B}\frac{\sin\omega_-(t-s)}{M\omega_-}\right],\\
\end{split}
\ee
where $\omega_-=\sqrt{\omega^2_m-2\omega_g^2}$. 
The $\langle \hat x_{B}(t)\rangle_c$ can be given in a similar way, where the only difference is to add a minus sign to all the $\cos\omega_-(t-s)$ and $\sin\omega_-(t-s)$ terms. For simplicity, we only write down the form in the $Q_m\rightarrow\infty$ approximation. It is clear that the quantum trajectory of mirror A is driven by the projective measurement on both the two outgoing fields. Substituting Eq.\,\eqref{eq:xAexpectation} into Eq.\eqref{eq:a2AB}, it is easy to see that there will be a correlation between $\tilde{a}_{2A}$ and $\tilde{a}_{2B}$ since $\tilde{a}_{2A/B}$ is also depend on $dW_{B/A}$. Moreover, there is also correlation if we measure two different outgoing field quadratures $\tilde{a}_{1A}$ and $\tilde{a}_{2B}$. This measurement-induced correlation is different from the quantum optical correlation that was studied in\,\cite{Miao2020} when the gravity is quantum, since classical gravity can not convey quantum information.  

$\bullet$ \textbf{Measurement of $(\tilde{a}_{2A},\tilde{a}_{2B})$---}
The optical correlation spectrum induced by the quantum measurement of $\tilde{a}_{2A}$, $\tilde{a}_{2B}$ is (we keep the leading term $\propto\alpha^4$):
\be\label{eq:spectrum_sn_a2b2}
\begin{split}
S^{\rm SN}_{a_{A2}a_{B2}}(\Omega)=\frac{2\hbar^2\alpha^4\omega_g^2}{M^2}&|\chi_m(\Omega)\chi_g(\Omega)|^2
\frac{\omega_q^2-\Omega^2}{1+\sqrt{1+\Lambda^4}}\\
&\times\bigg[\frac{\Lambda^4}{1+\sqrt{1+\Lambda^4}}+\frac{2\Omega^2}{\omega_q^2}\bigg],
\end{split}
\ee
where $\chi^{-1}_m(\Omega)=-\Omega^2+\omega_m^2-i\Omega\gamma_m$, $\chi^{-1}_g(\Omega)=-\Omega^2+\omega_-^2-i\Omega\gamma_m$ and $\Lambda=\sqrt{\hbar\alpha^2/M\omega_q^2}$. Later on, we will characterise the correlation between the two outgoing fields by the correlation level defined in Eq.\,\eqref{eq:correlation_level_22}\,\eqref{eq:correlation_level_12}, which also depends on the outgoing spectrum.  For example, the spectrum of the phase of the outgoing field $\tilde{a}_{A}$: $S_{a_{A2}a_{A2}}(\Omega)$ when we measure $(\tilde{a}_{2A},\tilde{a}_{2B})$:
\be\label{eq:spectrum_sn_a2a2}
\begin{split}
&S_{a_{A2}a_{A2}}(\Omega)\approx 1+\frac{\hbar^2\alpha^4}{M^2}|\chi_m(\Omega)\chi_g(\Omega)|^2\times\\
&\frac{\omega_g^4+(\Omega^2-\omega_q^2)^2+\gamma_m^2(\Omega^2+\omega_q^2)/2}{1+\sqrt{1+\Lambda^4}}
\bigg[\frac{\Lambda^4}{1+\sqrt{1+\Lambda^4}}+\frac{2\Omega^2}{\omega_q^2}\bigg],
\end{split}
\ee
where we also have $S_{a_{A2}a_{A2}}(\Omega)=S_{a_{B2}a_{B2}}(\Omega)$ because of symmetry.

\begin{figure}[h]
\centering
\includegraphics[width=0.5\textwidth]{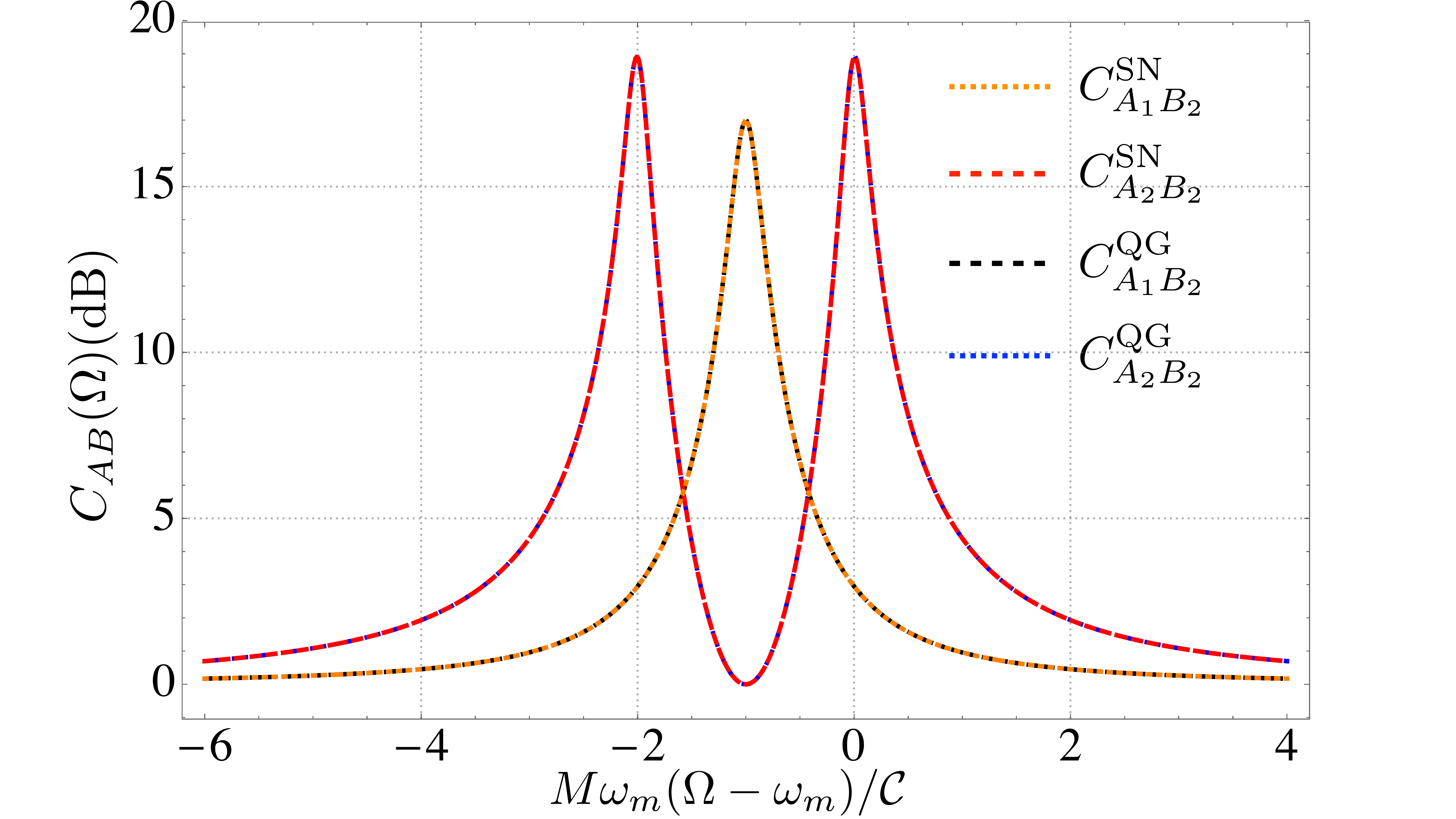}
\caption{The correlation spectrum between the output optical fields from the left and right cavities induced by the classical gravity, where the effect of thermal noise has been considered. It is easy to see that there is almost no difference between the quantum gravity case and the classical gravity case under the causal-conditional prescription.}\label{fig:SNcorrelation}
\end{figure}

To characterise the quantum measurement induced correlation, we define the correlation level at each frequency $\Omega$ to be (in terms of dB) :
\be\label{eq:correlation_level_22}
C^{\rm SN}_{A_2B_2}(\Omega)=-10\,{\rm Log}_{10}\left[1-\frac{|S^{\rm SN}_{a_{A2}a_{B2}}(\Omega)|^2}{S^{\rm SN}_{a_{A2}a_{A2}}(\Omega)S^{\rm SN}_{a_{B2}a_{B2}}(\Omega)}\right].
\ee
With Eq.\,\eqref{eq:spectrum_sn_a2b2}\eqref{eq:spectrum_sn_a2a2}, the terms in the square bracket of the above Eq.\,\eqref{eq:correlation_level_22} is approximately equal to:
\be
\begin{split}
C^{\rm SN}_{A_2B_2}&(\Omega)\approx-10\,{\rm Log}_{10}\\
&\left[
1-\left(\frac{2\omega_g^2(\omega_q^2-\Omega^2)}{\omega_g^4+(\omega_q^2-\Omega^2)^2+\gamma_m^2(\Omega^2+\omega_q^2)/2}\right)^2\right].
\end{split}
\ee
This approximate formula shows that there will be a strong correlation at $\Omega=(\omega_m,\omega_-)$, while zero correlation at $\Omega=\omega_q$, which well fits the peak features shown in Fig.\,\ref{fig:SNcorrelation} calculated using the exact formula.

The difference between the SN theory under causal-conditional prescription and the quantum gravity can be approximately represented as (the exact formula is a bit cumbersome):
\be
\begin{split}
\Delta_{\rm SN-QG}\bigg[\frac{S^{\rm cond}_{A_2A_2}-S_{A_2A_2}}{S_{A_2A_2}}\bigg]\approx
\left[\frac{\tilde{\omega}^2}{\tilde{\omega}^4+4(\omega_m/\omega_g)^2}\right]^3\frac{4\gamma^2_m}{\omega_g^2},
\end{split}
\ee
where $\tilde{\omega}^2\equiv(\Omega^2-\omega_q^2)/\omega_g^2$ and $S^{\rm cond}_{A_2A_2}$ is the conditional variance of the outgoing field $\tilde{a}_2$ when the $\tilde{b}_2$ is measured. This difference is negligibly small.

$\bullet$ \textbf{Measurement of $(\tilde{a}_{1A},\tilde{a}_{2B})$---}
The optical correlation spectrum induced by the quantum measurement of $\tilde{a}_{2A}$, $\tilde{a}_{2B}$ can be derived as (we keep the leading term $\propto\alpha^2$):
\be\label{eq:spectrum_sn_a1b2}
S^{\rm SN}_{a_{A1}a_{B2}}(\Omega)=S^{\rm QG}_{a_{A1}a_{B2}}(\Omega)\approx\frac{\hbar\alpha^2\omega_g^2}{M}\chi_m(\Omega)\chi_g(\Omega).
\ee
Note that, in this case, the difference between $S^{\rm SN}_{a_{A1}a_{B2}}(\Omega)$ and $S^{\rm QG}_{a_{A1}a_{B2}}(\Omega)$ is precisely zero. This can be understood as follows: from Eq.\,\eqref{eq:linear_cavity_trajectory}\eqref{eq:G-matrix}\eqref{eq:V-matrix}, one can see that the only term that proportional to $dW_B$ in the quantum trajectory of mirror-A has the coefficient equal to $\hbar/2$ when $\theta_A=0$ and $\theta_B=\pi/2$, which is independent from the second order correlation matrix $\mathbb{V}_{A/B}$. However, the difference between the SN theory under causal conditional prescription and the quantum gravity theory manifests in the $\mathbb{V}_{A/B}$. Therefore, the correlation between the $(\hat a_{A1}, \hat a_{B2})$ has no difference under SN theory comparing to the quantum gravity theory. In this case, the difference of the correlation level will be dominated by the difference of the outgoing spectrum.

The outgoing spectrum in the SN theory when we measure the $(\tilde{a}_{1A},\tilde{a}_{2B})$ is:
\be\label{eq:spectrum_sn_a2b2_b}
\begin{split}
&S^{\rm SN}_{a_{B2}a_{B2}}(\Omega)\approx1+\frac{\hbar^2\alpha^4}{M^2}|\chi_m(\Omega)\chi_g(\Omega)|^2\bigg[\omega_g^4+\\
&\left.\frac{(\omega_q^2-\Omega^2)^2+\gamma_m^2(\Omega^2+\omega_q^2)/2}{1+\sqrt{1+\Lambda^4}}\left(\frac{\Lambda^4}{1+\sqrt{1+\Lambda^4}}+\frac{2\Omega^2}{\omega_q^2}\right)\right],
\end{split}
\ee

Similarly, supposing the outgoing field quadratures $\tilde{a}_{1A}$ and $\tilde{a}_{2B}$ are measured, the correlation level can be defined as:
\be\label{eq:correlation_level_12}
C^{\rm SN}_{A_1B_2}(\Omega)=-10\,{\rm Log}_{10}\left[1-\frac{|S^{\rm SN}_{a_{A1}a_{B2}}(\Omega)|^2}{S_{a_{A1}a_{A1}}(\Omega)S_{a_{B2}a_{B2}}(\Omega)}\right],
\ee
where $S_{a_{A1}a_{A1}}(\Omega)=1$ since the outgoing amplitude quadrature does not carry the information of mirror displacement. With Eq.\,\eqref{eq:spectrum_sn_a1b2} and \eqref{eq:spectrum_sn_a2b2_b}, the correlation level can be approximately written as:
\begin{widetext}
\be
\begin{split}
C^{\rm SN}_{A_1B_2}(\Omega)=-10\,{\rm Log}_{10}\bigg[1-\bigg(1+
\frac{(\omega_q^2-\Omega^2)^2+\gamma_m^2(\Omega^2+\omega_q^2)/2}{\omega_g^4(1+\sqrt{1+\Lambda^4})}\left[\frac{\Lambda^4}{1+\sqrt{1+\Lambda^4}}+\frac{2\Omega^2}{\omega_q^2}\right]\bigg)^{-1}\bigg],
\end{split}
\ee
\end{widetext}
which demonstrates that the correlation level reaches maximum value when $\Omega=\omega_q$ as exhibits in Fig.\,\ref{fig:SNcorrelation}. We can also derive the difference between the SN theory and the quantum gravity when we measure the $(\tilde{a}_{1A},\tilde{a}_{2B})$, for $\Lambda\gg1$, the result has a simple form:
\be
\begin{split}
\Delta_{\rm SN-QG}\bigg[\frac{S^{\rm cond}_{A_2A_2}-S_{A_2A_2}}{S_{A_2A_2}}\bigg]\approx
\frac{1}{2}\frac{\tilde\omega^2}{\tilde\omega^4+1}\frac{\gamma_m^2}{\omega_g^2}\ll1,
\end{split}
\ee
while the result for $\Lambda\ll1$ has a complicated form but also a negligible magnitude. 

The above analysis shows that, the classical gravity induced correlation spectrum in the SN theory is almost indistinguishable from the quantum gravity induced correlation spectrum, under the causal-conditional prescription. Furthermore, considering the effect of thermal noise and finite quality factor $Q_m$, we plot the correlation spectrum  $C^{\rm SN}_{A_2B_2}(\Omega)$ and $C^{\rm SN}_{A_1B_2}(\Omega)$  in Fig.\,\ref{fig:SNcorrelation} using the sample parameters listed in Tab.\,\ref{tab:mutual_gravity}. The thermal noise enhances the indistinguishability of the SN theory from the quantum gravity. Under the pre-selection prescription, there will be no correlation of the output light fields when the gravity field is classical. For completeness, the correlation spectrum under the post-selection prescription is shown in the Supplementary Material, which is certainly different from that in quantum gravity theory. %In case the mutual gravity effect becomes stronger, the difference between the quantum gravity and classical gravity phenomenology will be distinguishable as we shown in Fig.\,\ref{fig:SNcorrelation}.

\begin{figure}[h!]
\centering
\includegraphics[width=0.5\textwidth]{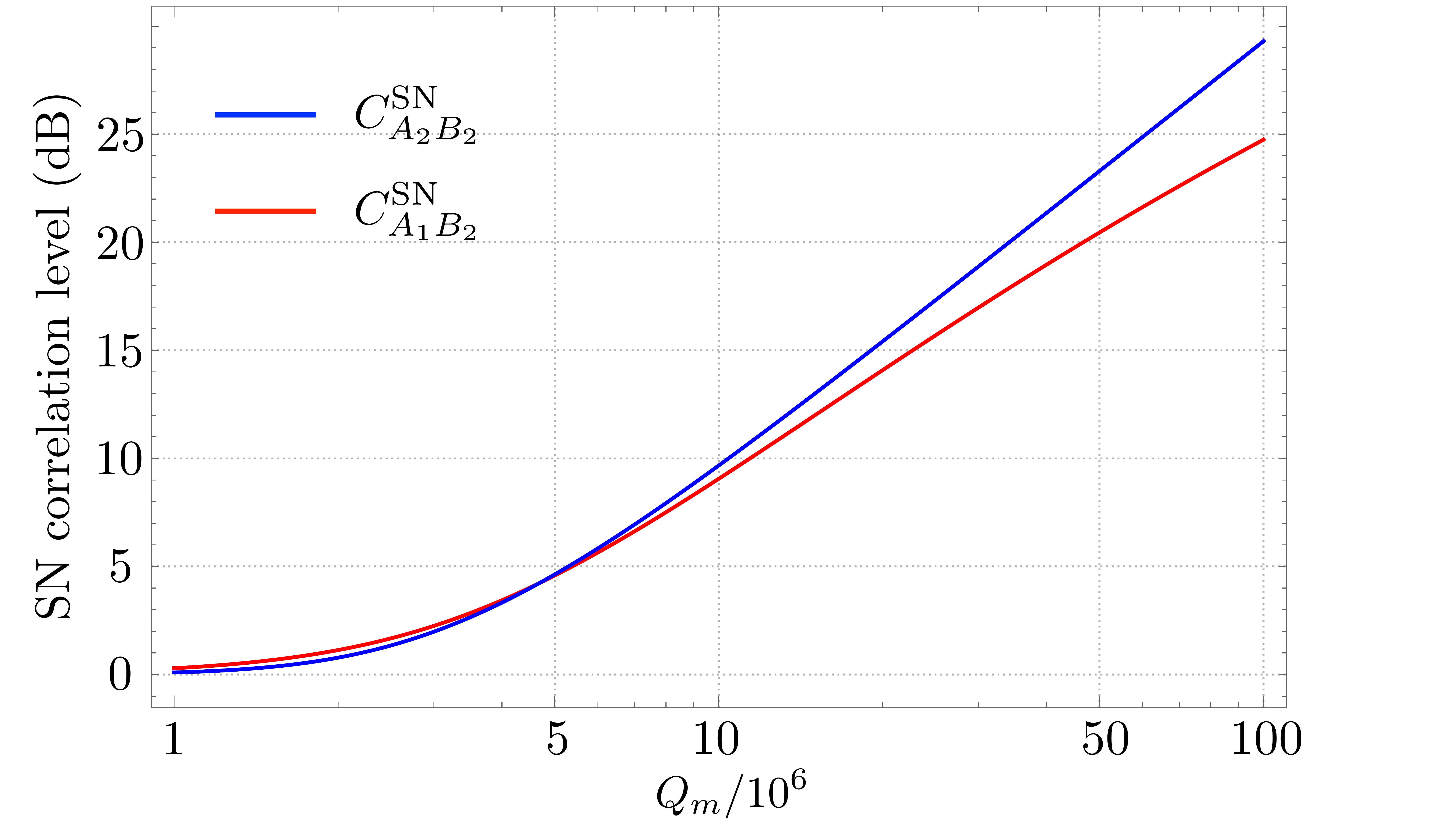}
\includegraphics[width=0.5\textwidth]{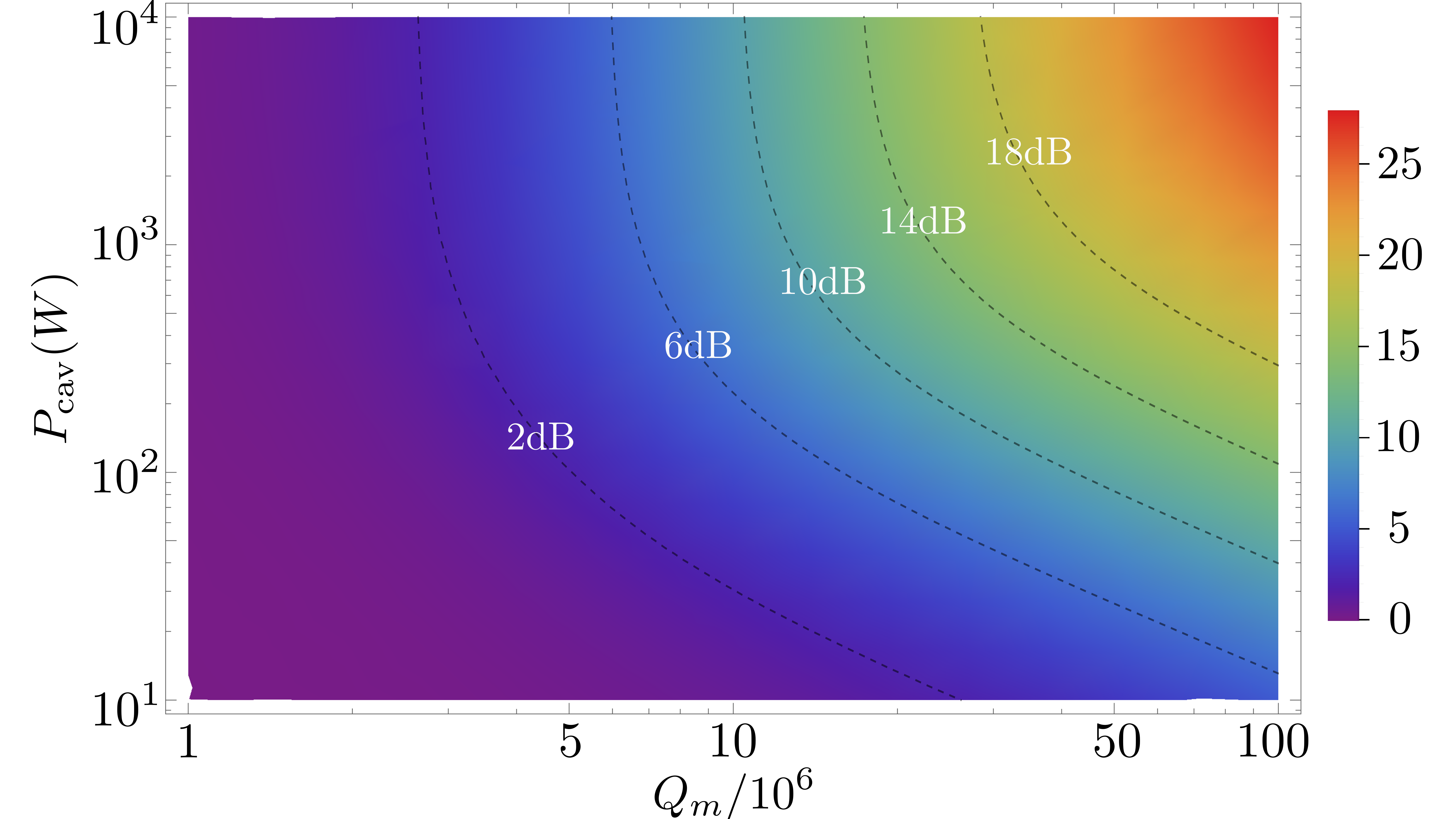}
\includegraphics[width=0.48\textwidth]{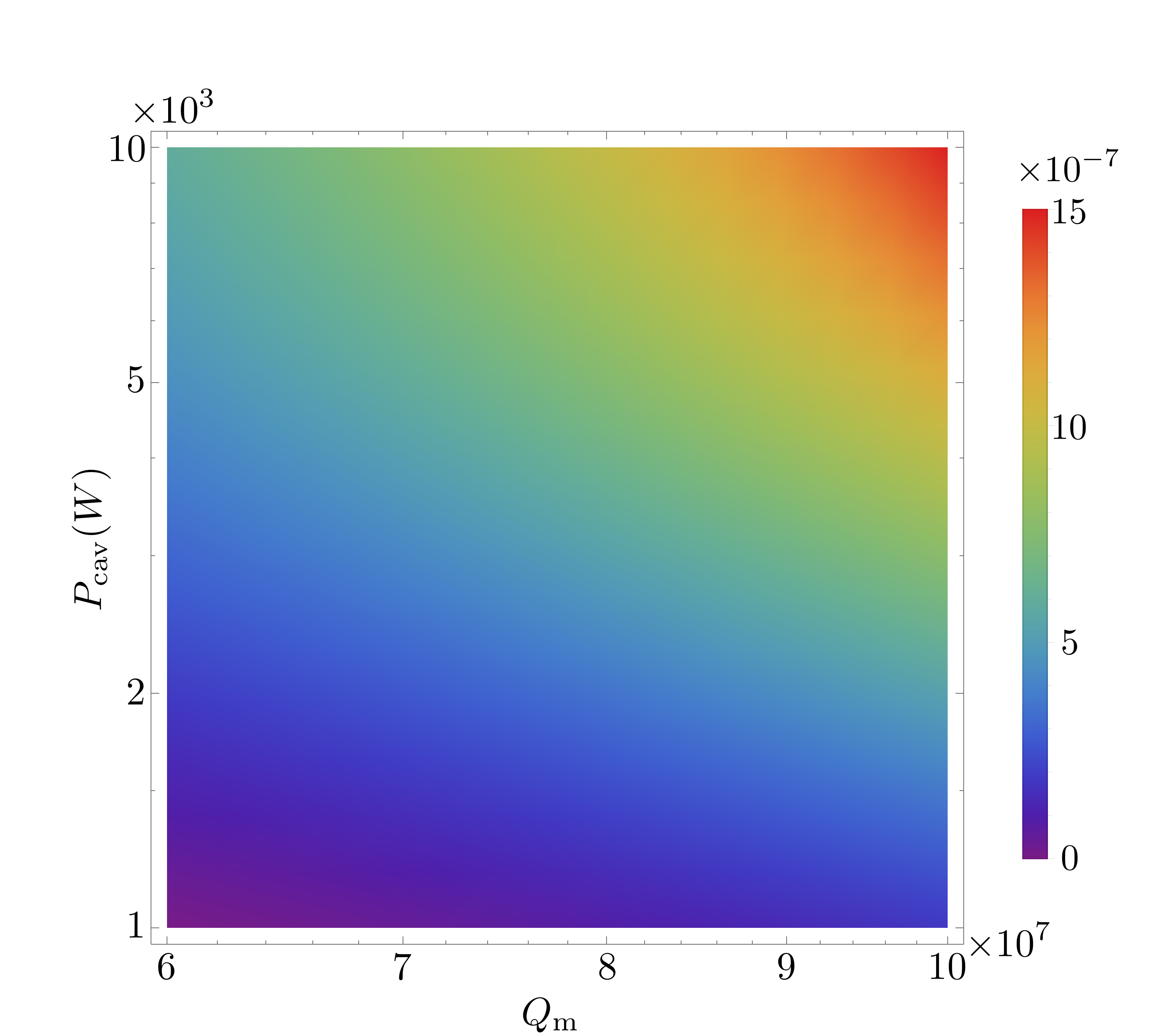}
\caption{The correlation between the output optical fields from the left and right cavities induced by the classical gravity, where the effect of thermal noise has been considered. Upper panel: the dependence of the maximum correlation value on the mechanical quality factor. Middle panel: the  dependence of the correlation on the mechanical quality factor $Q_m$ and the intracavity power $P_{\rm cav}$. It is almost identical to the result predicted by quantum gravity. Lower panel: The (almost zero) difference between the correlation level in SN theory and that in quantum gravity theory. We only choose a sub-region of parameter space in the middle panel with the difference is around $10^{-6}-10^{-7}$\,dB, while the value in other region of parameter space is much smaller.}\label{fig:SNcorrelationomegam}
\end{figure}

\section{The correlation and entanglement in nonlinear quantum mechanics}\label{sec.5}
The above discussion on the linear cavity systems shows a correlation of the outgoing light fields, which is similar to the entanglement in the standard quantum mechanics.  Looking into this phenomenon, this section devotes itself to a discussion of this correlation at the conceptual level.

Suppose we have two quantum systems A and B at an initial product state $|\Psi\rangle_i=|\Psi\rangle_A\otimes|\Psi\rangle_B$.  In the standard quantum mechanics without interaction between A and B, the state evolution follows:
\be
|\Psi\rangle_i\rightarrow \hat U^{AB}|\Psi\rangle_A\otimes|\Psi\rangle_B=\hat U^A|\Psi\rangle_A\otimes\hat U^B|\Psi\rangle_B,
\ee
where we have $\hat U^{AB}=\hat U^{A}\otimes \hat U^{B}$. The final state is still a product state without entanglement.  When these systems are coupled in the standard quantum mechanical way (say we have $\hat H=\hat H_A+\hat H_B+\hat H_{AB}$), the $\hat U^{AB}$ is not separable and the final state will be an entangled state:
\be
|\Psi\rangle_i\rightarrow \hat U^{AB}|\Psi\rangle_A\otimes|\Psi\rangle_B=\sum_i c_i|\Psi\rangle_{Ai}\otimes|\Psi\rangle_{Bi}.
\ee
This formula means that the projective measurement on $|\Psi\rangle_{Ai}$ will immediately collapse the joint quantum state onto $|\Psi\rangle_{Bi}$, which exhibits a correlation between the measurement result of systems A and B.  A system consisting of two optomechanical devices coupled via quantum gravity is in this category.

Moreover, if we have a nonlinear quantum mechanics such as Schroedinger-Newton theory, the state evolution follows:
\be\label{eq:stateSN}
\begin{split}
|\Psi\rangle_i\rightarrow &\hat U^{AB}|\Psi_A\rangle\otimes|\Psi_B\rangle=\\
&\hat U^{A}(|\Psi_B\rangle)|\Psi_{A}\rangle\otimes\hat U^{B}(|\Psi_A\rangle)|\Psi_{B}\rangle,
\end{split}
\ee
where we have used the example Hamiltonian
\be
\hat H=\hat H_A+\hat H_B+\hat H^A_{\rm int}(|\Psi_B\rangle)+\hat H^B_{\rm int}(|\Psi_A\rangle),
\ee
with
\be
\begin{split}
\hat U^{AB}=&{\rm exp}\left[-\frac{i}{\hbar}[\hat H_A+\hat H^A_{\rm int}(|\Psi_B\rangle)]t\right]\otimes\\
&{\rm exp}\left[-\frac{i}{\hbar}[\hat H_B+\hat H^B_{\rm int}(|\Psi_A\rangle)]t\right].
\end{split}
\ee
The final state is still a product state, however only in mathematical appearance. In reality, the quantum states of A and B are correlated subtly as shown in Eq.\,\eqref{eq:stateSN}: measurement on the system state $|\Psi_{A/B}\rangle$ would induce the change of evolution operator $\hat U^{B/A}(|\Psi_{A/B}\rangle)$, thereby affecting the evolution of $|\Psi_{B/A}\rangle$. Therefore, although Eq.\,\eqref{eq:stateSN} has the mathematical form of a pure product state, there is still a correlation between the system A and B, which only exists in the nonlinear quantum mechanics such as the SN theory.

Under the pre-selection prescription, since we do not measure the initial state in a real experiment, we have the interaction Hamiltonian as $\hat H^{B/A}_{\rm int}(|\Psi_{A/B}(t=0)\rangle)$. Therefore, this means the final states $|\Psi_{A/B}(t_f)\rangle$ do not depend on each other thereby existing no correlation. However, under the post-selection or causal-conditional prescription, the interaction Hamiltonian is $\hat H^{B/A}_{\rm int}(|\Psi_{A/B}(t_f)\rangle)$ or $\hat H^{B/A}_{\rm int}(|\Psi_{A/B}(t_f)\rangle_c)$ ($|\Psi_{A/B}(t_f)\rangle_c$ is the conditional final state generated by continuous quantum measurement), respectively. Therefore there will be correlations between the two systems due to the above-discussed reasons. Further investigation of the correlations in nonlinear quantum mechanics is beyond the scope of this work and will be written elsewhere.

\section{Discussion and Summary}\label{sec.6}
Testing the gravitational law in the quantum era now becomes a blooming field, where many proposals were raised in recently years\,\cite{Snowmass2022,Pedernales2022,Matsumura2022,Chevalier2020,Matsumura2020,Liu2021,Pikovski2012}. These proposals covered many different aspects and the phenomenologies in the quantum/gravity interface, and triggered many discussions and even debates on these phenomenologies\,\cite{Pikovski2015,Adler2016,Bonder2015,Pang2016}. This work devotes a deeper understanding of the phenomenologies of Schroedinger-Newton theory in the quantum optomechanical system, which is motivated by the theory of semi-classical gravity. We pointed out that the nonlinear term in the Schroedinger-Newton equation breaks the time-symmetry of quantum measurement in the standard quantum mechanics and brings additional complexity. We specifically analysed the Schroedinger-Newton phenomenology under the causal-conditional prescription by establishing the stochastic master equation in the Schroedinger picture. We apply the master equation to study the behaviour of the optomechanical systems exerted by the single mirror's self-gravity force and mutual gravity between the two mirrors, under the continuous quantum measurement. Our results show that, different from the predictions of the previous work\,\cite{Miao2020,Datta_2021,Helou2017,Yang2013}, the semi-classical gravity effect under the causal-conditional prescription is very difficult to be distinguished from the quantum gravity effect with ponderomotive squeezing or correlation/spectrum of outgoing fields, 
even in the case of optomechanical system exerted by the single mirror's (relatively strong) classical self-gravity when we considered the thermal environment. The previously predicted feature\,\cite{Helou2017} at $\omega_q=\sqrt{\omega_m^2+\omega_{\rm SN}^2}$ diminishes, mainly because that the continuous quantum measurement induces the collapse of the joint mirror-light wave function and creates a stochastic quantum trajectory of the mirror state. This quantum trajectory can also participate in the classical gravitational interaction process and create correlations, as we have discussed in Section\,\ref{sec.4}. Since the causal-conditional prescription fits our intuition about the continuous quantum measurement process, the phenomenology obtained in this work is an important reference for the experiment: the phenomena observed using the methods proposed in\,\cite{Yang2013,Helou2017,Miao2020,Datta_2021} could not be the sufficient condition to recognize quantum gravity or rule out SN theory, under the current experimental state-of-arts and the weak  gravitational interactions. It is possible to test the quantum nature of gravity using the pondermotive squeezing spectrum of the optomechanical system exerted by the mirror's self-gravity only if the environmental temperature takes extremely low values.

Another point that worthy to be discussed is the semi-classical gravity itself. Semi-classical gravity is usually criticized since it has contradictions to the many-worlds interpretations\,\cite{Everett1957}, and some inconsistencies when combining a quantum world with a classical space-time theater\,\cite{Page1981,Carlip_2008}. However, as Steven Carlip pointed out\,\cite{Carlip_2008}, \emph{``theoretical arguments aganist such mixed classical-quantum models are strong, but not conclusive, and the question is ultimately one for experiment."}. In particular, the strong argument by Page and Geilker on the contradiction between semi-classical gravity and the many-world interpretation may diminish if the wavefunction collapse can be explained within the quantum mechanics, which is still an open question. As a side-remark, Stamp et.al recently proposed an alternative approach for reconciling quantum mechanics and gravity, which is called correlated-worldline\,(CWL) theory\,\cite{Stamp2018,Barvinsky2021,Stamp2022}. The CWL theory is fundamentally a quantum gravity theory, of which the feature is that the different paths in the path-integral are correlated via gravity. In the infra-limit, the CWL theory will reduce to the Schroedinger-Newton theory\,\footnote{Private communications with Philip C. E. Stamp.}, which will be discussed elsewhere. Therefore, pursuing the experimental/theoretical research in testing the quantumness of gravity is still very important, despite those criticism of semi-classical gravity. 

%\R{This is a general approach to all nonlinear quantum mechanics---}

\acknowledgements
We thank Animesh Data, Mikhail Korobko, Sergey Vyachatnin, Enping Zhou, Daiqin Su, Philip C. E. Stamp and Markus Aspelmeyer for helpful discussions.  Y. M. is supported by the start-up fund provided by Huazhong University of Science and Technology. H. M. is supported by State Key Laboratory of Low Dimensional Quantum Physics and the start-up fund from Tsinghua University. Y. C. is supported by Simons Foundation.

\bibliographystyle{unsrt}
\bibliography{causal-conditional}

\end{document}